\documentclass{aa}
\usepackage[varg]{txfonts}
\usepackage{natbib, amsmath, amssymb, amsfonts, graphicx}
\usepackage{mathtools}
\usepackage{subfig}
\usepackage[squaren, Gray, cdot]{SIunits}
\usepackage[citecolor=blue, linkcolor=blue, urlcolor = black, colorlinks = true]{hyperref}

\title{The traditional approximation of rotation for rapidly rotating stars and planets}
\subtitle{II. Deformation and differential rotation}
\author{H. Dhouib \inst{1}
\and V. Prat \inst{1}
\and T. Van Reeth \inst{2}
\and S. Mathis \inst{1}}

\institute{Département d’Astrophysique-AIM, CEA/DRF/IRFU, CNRS/INSU, Université Paris-Saclay, Université Paris-Diderot, Université
de Paris, F-91191 Gif-sur-Yvette, France\\\email{hachem.dhouib@cea.fr}
\and Institute of Astronomy, KU Leuven, Celestijnenlaan 200D, 3001 Leuven, Belgium} 

\titlerunning{The traditional approximation of rotation for rapidly rotating stars and planets. II.}
\authorrunning{H. Dhouib et al.}

\abstract
{We examine the dynamics of low-frequency gravito-inertial waves (GIWs) in differentially rotating deformed radiation zones in stars and planets by generalising the traditional approximation of rotation (TAR). The TAR treatment was built on the assumptions that the star is spherical (i.e. its centrifugal deformation is neglected) and uniformly rotating. However, it has been generalised in our previous work by including the effects of the centrifugal deformation using a non-perturbative approach. In the meantime, TAR has been generalised in spherical geometry to take the differential rotation into account.}
{We aim to carry out a new generalisation of the TAR treatment to account for the differential rotation and the strong centrifugal deformation simultaneously.}
{We generalise our previous work by taking into account the differential rotation in the derivation of our complete analytical formalism that allows the study of the dynamics of GIWs in  differentially and rapidly rotating stars.}
{We derived the complete set of equations that generalises the TAR, simultaneously taking the full centrifugal acceleration and the differential rotation into account. Within the validity domain of the TAR, we derived a generalised Laplace tidal equation for the horizontal eigenfunctions and asymptotic wave periods of the GIWs, which can be used  to probe the structure and dynamics of differentially rotating deformed stars with asteroseismology.}
{A new generalisation of the TAR, which simultaneously takes into account the differential rotation and the centrifugal acceleration in a non-perturbative way, was derived. This generalisation allowed us to study the detectability and the signature of the differential rotation on GIWs in rapidly rotating deformed stars and planets. We found that the effects of the differential rotation in  early-type deformed stars on GIWs is theoretically largely detectable in modern space photometry using observations from \textit{Kepler} and TESS.}

\begin{document}
\maketitle
\section{Introduction}
 
Understanding how angular momentum and chemicals are transported in the interiors of stars (and planets) along their evolution is one of the key challenges of modern stellar (and planetary) astrophysics. Indeed, rotation modifies their structure, their chemical stratification, their internal flows and magnetism, and their mass losses and winds \citep[e.g.][and references therein]{Maeder2009,Mathis2013,aerts2019}. In this quest, asteroseismology has bought a fundamental breakthrough by demonstrating that all stars are the seat of a strong extraction of angular momentum during their evolution in comparison with the predictions by stellar models taking the rotation into account following the standard rotational transport and mixing theory \citep{Eggenberger2012,Marques2013,Ceillier2013,Cantiello2014,ouazzani2019}. This was first obtained thanks to mixed pulsation modes splitted by rotation propagating in evolved low- and intermediate-mass stars \citep{Beck2012RG,Mosser2012RG,Deheuvels2012RG,Beck2014RG,Deheuvels2014RG,Deheuvels2015RG,DiMauro2016RG,Triana2017RG,Gehan2018RG,Beck2018RG,Tayar2019RG,Deheuvels2020RG}. Then, observations of oscillation
%acoustic, gravity, and gravito-inertial\footnote{i.e. gravity modes modified by rotation when it cannot be treated as a perturbation.}
modes in F- and A-type stars \citep{Kurtz2014A,Saio2015A,Beddingetal2015Gamma,Keen2015Gammabin,vanreeth2015,vanreeth2016,Schmid2016F,Murphy2016Gamma,Sowicka2017F,Guoetal2017Fbin,vanreeth2018,Saio2018Gamma,Mombarg2019F,Li2019Gamma,Li2020Gamma,Ouazzani2020Gamma,Saio2021Gamma} and in B-type stars \citep{Papics2015B,Triana2015B,Moravveji2016,Papics2017,Kallingeretal2017Bbin,Buysschaert2018B,Szewczuk2018,Pedersen2021B,Szewczuk2021B} provided us new Rosetta stones to constrain the transport of angular momentum in the whole Hertzsprung-Russell diagram. More particularly, this pushes gravity- and gravito-inertial mode pulsators such as $\gamma$-Doradus and SPB stars at the forefront of this research. For instance, recent theoretical developments have demonstrated how it is possible to probe stellar internal rotation in $\gamma$-Doradus stars from their surface to their convective core \citep[][]{Ouazzani2020Gamma,Saio2021Gamma}. These stars are rapid rotators for a large proportion of them. Therefore, it is necessary to study gravito-inertial modes. These modes are gravity modes, which propagate only in stably stratified stellar radiation zones under the action of the restoring buoyancy force in the absence of rotation, which are modified by rotation. If their frequency is super-inertial (i.e. above the inertial frequency $2\Omega$, $\Omega$ being the stellar angular velocity), they are propagating in stellar radiation zones and evanescent in convective regions. If their frequency is sub-inertial (below $2\Omega$) they propagate in an equatorial belt in radiation zones and they become propagative inertial waves in convective regions \citep[e.g.][]{Dintrans2000,MNT2014}. The challenge of studying these waves is that the equation describing their dynamics are intrinsically bi-dimensional and non-separable \citep{Dintrans1999,Prat2016,Mirouh2016,Prat2018}. This makes the development of seismic diagnosis difficult analytically \citep[]{prat2017} or expansive in computation time when using 2D oscillation and stellar structure codes \citep[e.g.][]{ouazzani2017,Reese2021} in the general case.

 In this framework, the traditional approximation of rotation (TAR), which has been first introduced in geophysics \citep{eckart1960} to treat the propagation of gravito-inertial waves (GIWs) in the case where the Coriolis acceleration can be neglected in front of the buoyancy force in the direction of the entropy and chemical stable stratifications, is very useful. Indeed, it allows one to consider that the velocity of gravito-inertial waves are mostly horizontal and to neglect the latitudinal component of the rotation vector in the momentum equation. That leads to decoupling the vertical and horizontal directions by keeping the non-perturbative action of the Coriolis acceleration in the latitudinal and azimuthal directions and neglecting it along the vertical one. Propagation equations become separable as in the non-rotating case \citep[e.g.][]{Bildsten1996,lee+saio1997}. In addition, the derivation of powerful and flexible seismic diagnostics using the period spacing between consecutive high radial order gravito-inertial modes becomes possible in uniformly rotating spherical stars \citep{Bouabid2013}. These period spacing provide constrains on properties of the chemical stratification and the rotation rate (through their slope) near the interface between the convective core and the radiative envelope in rapidly rotating intermediate-mass $\gamma$-Doradus \citep[e.g.][]{vanreeth2015,vanreeth2016,vanreeth2018,ouazzani2017,Christophe2018,Li2019Gamma,Li2020Gamma,Saio2021Gamma} and SPB stars \citep[e.g.][]{Papics2015B,Papics2017,Moravveji2016,Szewczuk2018,Pedersen2021B}, thanks to the high precision of space-based photometric observations \citep[e.g.][and references therein]{aerts2019,aerts2021}. This allows us to build intensive forward modelling of g-mode pulsating stars to constrain their internal structural and dynamical properties.\\

This approximation, in its standard version, is only applicable for low-frequency GIWs propagating in strongly stably stratified zones of uniformly rotating spherical stars. In this case a set of assumptions should be verified: (i) the buoyancy force is stronger than the Coriolis acceleration (i.e. $ 2 \Omega \ll N $, where $N$ is the Brunt-Väisälä frequency)  in the direction of stable entropy or chemical stratification, (ii) the Brunt-Väisälä frequency is larger than the frequency of the waves in the rotating frame $\omega$ ($ \omega \ll N $), (iii) the rotation is assumed to be uniform and (iv) the star is assumed to be spherical, in other words, the centrifugal deformation of the star is neglected (i.e. $\Omega \ll \Omega_{\mathrm{K}} \equiv \sqrt{G M / R^{3}}$,  where $\Omega_{\mathrm{K}}$ is the Keplerian critical (breakup) angular velocity, and $G$, $M$, and $R$ are  the universal constant of gravity, the mass of the star, and the stellar radius, respectively).

Recently, the TAR has been examined in deformed stars. First, \cite{mathis+prat2019} have included the centrifugal acceleration for slightly deformed stars using a first-order perturbative approach. Then, this perturbative framework has been optimised for practical applications to one-dimensional stellar structure models by \cite{Henneco2021}. Second, in \citet[][hereafter Paper I]{dhouib2021}, we have used a non-perturbative approach to include the centrifugal acceleration for strongly deformed stars and planets by studying the dynamics of low-frequency GIWs in a general spheroidal coordinate system defined by \cite{bonazzola1998} which follows the shape of a deformed star. It is important to note that in these studies the rotation was assumed to be uniform. These semi-analytical studies demonstrated that centrifugal effects' signatures are potentially detectable. Moreover, their results are in good qualitative agreement with direct computations of gravito-inertial modes in centrifugally deformed stellar models using 2D oscillation codes that take into account the full Coriolis and centrifugal accelerations such as the Toulouse Oscillation Program (TOP) \citep{Ballot2010,Ballot2012} and the adiabatic code of oscillation including rotation (ACOR) code \citep{Ouazzani2012,Ouazzani2015,ouazzani2017}. In the case where the centrifugal acceleration is treated as a perturbation, but where the full Coriolis acceleration was taken into account, we can refer to the Tohoku oscillation code \citep{lee+saio1987,lee+baraffe1995}.
The chosen coordinate system is used in ACOR, TOP and in the Evolution STEllaire en Rotation (ESTER) code \citep{espinosa+rieutord2013} that computes the structure and the stationary internal hydrodynamics of (differentially-)rotating early-type stars such as the observed gravito-inertial modes pulsators we are studying.

However, the solid-body rotation assumption has to be potentially abandoned along the evolution of real stars (and planets) where gradients of angular velocity can develop, both in the radial and in the latitudinal directions. They can be triggered by the redistribution of angular momentum induced by stellar winds \citep[e.g.][]{zahn1992}, structural adjustments \citep{Maeder&Meynet2000, Decressin2009}, and by transport processes \citep{Mathis2013}. First, as shown in \cite{charbonnel2013} and \cite{Gallet+Bouvier2015},  early main sequence low-mass stars are potentially subject to a strong differential rotation. Moreover, \cite{aerts2019} and references therein (we refer the reader to those provided in the first paragraph for F, A, and B-type stars) pointed out that the radiative envelope of intermediate-mass stars are the seat of a weak differential rotation for those observed with a small difference of rotation between the stellar surface and the core. This is also predicted by numerical simulations and models of different transport mechanisms like internal gravity waves \citep {Rogers2013,rogers2015} or meridional flows \citep{Maeder&Meynet2000,Decressin2009,Rieutord2016,ouazzani2019}. In this framework, \citet{ogilvie+lin2004} and \citet{mathis2009} have generalised the TAR by including the effects of general differential rotation on low-frequency GIWs propagating in spherical stars (and planets). This new formalism has been successfully applied to gravito-inertial mode pulsators observed by the \textit{Kepler} \citep{borucki2009} space telescope. Indeed, \citet{vanreeth2018} have applied
it to derive the variation of the asymptotic period spacing in the case of a weak radial differential rotation. They evaluated the sensitivity of GIWs period spacing to the effect of such a radial shear in spherical stars and they demonstrated that they can be potentially detected if observing several modes. 

Since both effects of the centrifugal deformation and differential rotation are potentially observable, we generalise here these previous works, treating the case of a general  differential rotation in deformed stars (and planets). 
We begin in Sect.\,\ref{sect:Hydrodynamic_equations} with the derivation of the system of linearised hydrodynamic equations for GIWs in spheroidal coordinates which follows the shape of a differentially rotating deformed body. We choose again the coordinate system that was introduced by \cite{bonazzola1998} which is used in 2D numerical models of rotating stars and pulsation codes. In Sect.\,\ref{sect:Generalised_TAR}, we build the generalised TAR in this set of equations by adopting the adequate assumptions. In Sect.\,\ref{sect:Dynamics_GIWs}, we rewrite the oscillation equations in the form of a generalised Laplace tidal equation and deduce the asymptotic frequencies of low-frequency GIWs. In Sect.\,\ref{sect:results}, we carry out a numerical exploration of the eigenvalues and Hough eigenfunctions of the generalised Laplace tidal equation within the domain of validity of the TAR. We model early-type stars with 2D ESTER models \citep{espinosa+rieutord2013}. In Sect.\,\ref{sect:seismic_diagnosis}, we quantify the differential rotation and the centrifugal deformation combined effects on the asymptotic period spacing pattern and we discuss their detectability. Finally, we discuss and summarise our work and results in Sect.\,\ref{sect:conclusion}.

\section{Hydrodynamic equations in differentially rotating deformed stars }\label{sect:Hydrodynamic_equations}
\subsection{Spheroidal geometry}
\begin{figure}
    \centering
    \resizebox{\hsize}{!}{\includegraphics{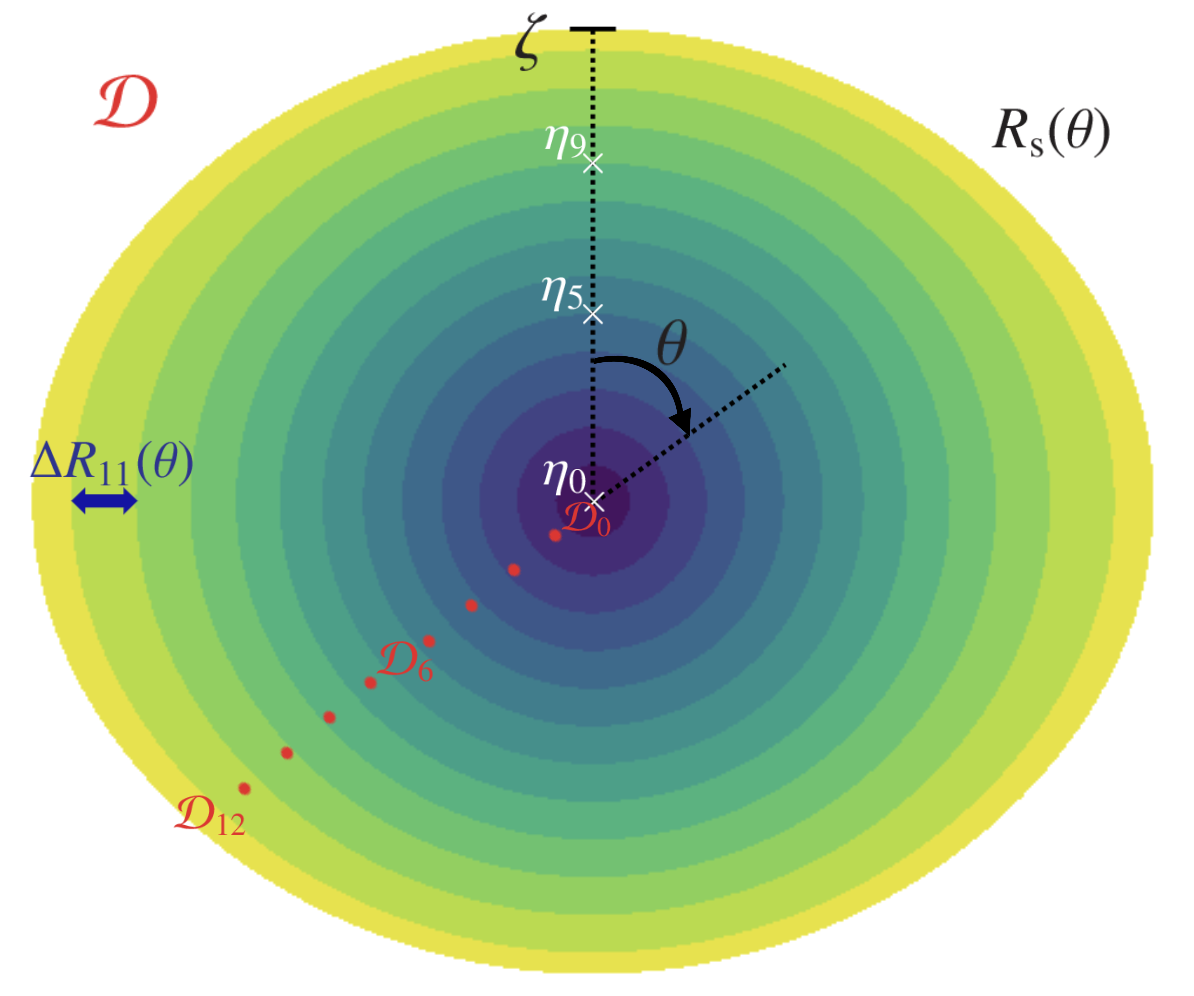}}
    \caption{Spheroidal coordinate system with $n=13$ subdomains used to compute the equilibrium model of a star rotating at $60\%$ of its break-up velocity corresponding to the case of the $3 \mathrm{M}_{\odot}$ ESTER model (with the central fraction in Hydrogen $X_{\rm c} = 0.7$) studied in Sect.\,\ref{sect:results}. $\zeta$ and $\theta$ are the pseudo-radius and the colatitude, respectively. $\eta_i$ and $\Delta R_i(\theta)$ are defined in our mapping (Eq.\;\ref{eq:mapping}).}
    \label{fig:spheroidal_geometry}
\end{figure}

As in Paper I, here, we use the spheroidal coordinate system $(\zeta, \theta, \varphi)$ proposed by \citet{bonazzola1998}, where $\zeta$ is the pseudo-radial coordinate, $\theta$ the colatitude and $\varphi$ the azimuth. Following \citet{espinosa+rieutord2013}, this new coordinate system, illustrated in Fig.\,\ref{fig:spheroidal_geometry}, can be linked to the usual spherical one $(r, \theta, \varphi)$ via the following mapping
\begin{equation}\label{eq:mapping}
    r(\zeta, \theta)=a_{i} \xi \Delta \eta_{i}+R_{i}(\theta)+A_{i}(\xi)\left(\Delta R_{i}(\theta)-a_{i} \Delta \eta_{i}\right), \quad \eta_{i} \leq \zeta \leq \eta_{i+1},
\end{equation}
in the $i$th subdomain $\mathcal{D}_{i \in  \llbracket0,n-1 \rrbracket} \in\left[R_{i}(\theta), R_{i+1}(\theta)\right] $ of the spheroidal domain $\mathcal{D}$ where $R_{i \in  \llbracket0,n \rrbracket}(\theta)$ are series of functions, such that $R_{n}(\theta)=R_{\rm s}(\theta)$ is the outer boundary and $R_{0}(\theta)=0$ is the centre. Additionally, $\eta_{i} =R_{i}(\theta=0)$ are the polar radii of the interfaces between the subdomains, $\Delta \eta_{i} =\eta_{i+1}-\eta_{i}$, $\Delta R_{i}(\theta) =R_{i+1}(\theta)-R_{i}(\theta)$ and $\xi =(\zeta-\eta_{i})/\Delta \eta_{i}$. The functions $A_{i \in  \llbracket1,n-1 \rrbracket}(\xi)=-2 \xi^{3}+3 \xi^{2}$ and $A_{0}(\xi)=-1.5 \xi^{5}+2.5 \xi^{3}$ and the constants $a_{i}=1$  are chosen to satisfy the boundary conditions between the different subdomains.
We recall also the spheroidal base $(\vec{a}_{\zeta}, \vec{a}_{\theta}, \vec{a} _ {\varphi})$ defined using the mapping (Eq.\;\ref{eq:mapping}) \citep{rieutord2005,lignieres2006,reese2006}
\begin{equation}\left\{\begin{array}{lcc} \begin{aligned}
\vec{a}_{\zeta} &=\frac{\zeta^{2}}{r^{2}} \vec{e}_{r} \\
\vec{a}_{\theta} &=\frac{\zeta}{r^{2} r_{\zeta}}\left(r_{\theta} \vec{e}_{r}+r \vec{e}_{\theta}\right) \\
\vec{a}_{\varphi} &=\frac{\zeta}{r r_{\zeta}} \vec{e}_{\varphi}
\end{aligned}\end{array}\right.,\end{equation}
where $r_\zeta\equiv\partial_\zeta r$, $r_\theta\equiv\partial_\theta r$, $(\vec{e}_{r}, \vec{e}_{\theta}, \vec{e} _ {\varphi})$ is the usual spherical base, $\varepsilon = 1-R_{\rm pol}/R_{\rm eq}$ is the flatness, $R_{\rm s}(\theta)$, $R_{\rm eq}$ and $R_{\rm pol}$ are the surface, the equatorial and the polar radii, respectively.

\subsection{Linearised hydrodynamic equations in  spheroidal coordinates}
To treat the wave dynamics in  differentially rotating, strongly deformed stars and planets, we derive the complete adiabatic inviscid system of equations in  spheroidal coordinates. 
First, the linearised momentum equation for an inviscid fluid is written as
\begin{multline} \label{eq:radial_momentum}
    (\partial_t+\Omega\partial_\varphi) \left[\frac{\zeta^2 r_{\zeta} }{r^{2}}v^{\zeta}+\frac{\zeta r_{\theta} }{r^{2}}v^{\theta}\right]
    \\=2 \Omega     \frac{\zeta \sin \theta }{r}v^{\varphi}-\frac{1}{\rho_{0}}\partial_{\zeta} P^{\prime}+\frac{\rho^{\prime}}{\rho_{0}^2} \partial_{\zeta} P_{0}- \partial_{\zeta} \Phi^{\prime},
\end{multline}

\begin{multline} \label{eq:latitudinal_momentum}
    (\partial_t+\Omega\partial_\varphi)  \left[\frac{\zeta^2 r_{\theta}}{r^{2}}v^{\zeta} + \frac{\zeta \left( r^{2}+r_{\theta}^{2} \right)}{r^{2} r_{\zeta}}v^{\theta}\right]\\
    =2 \Omega    \frac{\zeta\left(r_{\theta} \sin \theta+r \cos \theta \right) }{r r_{\zeta}}v^{\varphi}-\frac{1}{\rho_{0}}\partial_{\theta} P^{\prime}+\frac{\rho^{\prime}}{\rho_{0}^2} \partial_{\theta} P_{0}- \partial_{\theta} \Phi^{\prime},
\end{multline}

\begin{multline} \label{eq:azimuthal_momentum}
    \frac{\zeta }{r_{\zeta}}(\partial_t+\Omega\partial_\varphi) v^{\varphi}=- 2 \Omega \frac{ \zeta^{2} \sin \theta }{r} v^{\zeta}-2 \Omega \frac{ \zeta\left(r_{\theta} \sin \theta+r \cos \theta\right) }{r r_{\zeta}}v^{\theta}
    \\-\frac{1}{\rho_0 \sin \theta}\partial_{\varphi} P^{\prime}- \frac{\zeta \sin{\theta}}{r_\zeta}\left(\zeta \partial_\zeta \Omega u^\zeta + \partial_\theta \Omega u^\theta\right)
    -\frac{1}{\sin \theta} \partial_{\varphi} \Phi^{\prime},
\end{multline}
where $v^{\zeta}$, $v^{\theta}$, and $v^{\varphi}$ are the contravariant components of the velocity field $\vec{v}=v^{i} \vec{a}_{i}$ and $\vec{\Omega}=\Omega(\zeta, \theta)(\cos{\theta}\vec{e_r}-\sin{\theta}\vec{e_\theta})$ is the differential angular velocity of the star. $\rho$, $\Phi$ and $P$ are the fluid density, gravitational potential and pressure, respectively. Each of these scalar quantities has been expanded as
\begin{equation*}
X(r, \theta, \varphi, t)=X_0(r,\theta)+X^\prime(r, \theta, \varphi, t),
\end{equation*}
where $X_0$ is the hydrostatic component of $X$ and $X^\prime$ the wave's associated linear fluctuation.
Following \citet{cowling1941}, we neglect the fluctuation of the gravitational potential.

Next, the linearised continuity equation is obtained
\begin{multline} \label{eq:continuity}
    (\partial_t+\Omega\partial_\varphi)  \rho^{\prime}=-\frac{\zeta^{2} \partial_{\zeta} \rho_{0}}{r^{2} r_{\zeta}} v^{\zeta}-\frac{\zeta \partial_{\theta} \rho_{0}}{r^{2} r_{\zeta}} v^{\theta}
    \\-\frac{\zeta^{2} \rho_{0}}{r^{2} r_{\zeta}}\left[\frac{\partial_{\zeta}\left(\zeta^{2} v^{\zeta}\right)}{\zeta^{2}}+\frac{\partial_{\theta}\left(\sin \theta v^{\theta}\right)}{\zeta \sin \theta}+\frac{\partial_{\varphi} v^{\varphi}}{\zeta \sin \theta}\right].
\end{multline}
Then, the linearised energy equation in the adiabatic limit  is derived
\begin{equation}\label{eq:energy_eq}
    (\partial_t+\Omega\partial_\varphi) \left(\frac{1}{\Gamma_{1}}\frac{P^{\prime}}{P_0}- \frac{\rho^{\prime}}{\rho_0}\right)=\frac{ N^{2} }{\left\|\vec{g}_{\rm eff}\right\|^{2}} \vec{v} \cdot \vec{g}_{\rm eff},
\end{equation}
where $\Gamma_{1}=(\partial \ln P_0 / \partial \ln \rho_0)_{S}$ ($S$ being the macroscopic entropy) is the adiabatic exponent, $\vec{g}_{\rm eff}=-\vec{\nabla} \Phi_0+\frac{1}{2}\Omega^2\vec{\nabla}(r^2\sin^2{\theta})=\vec{\nabla}P_0/\rho_0$ is the background effective gravity which includes the centrifugal acceleration and $N^2$ the squared Brunt-Väisälä frequency (or buoyancy frequency) given by
\begin{equation}\label{eq:N2}
    N^{2}(\zeta, \theta)=\vec{g}_{\rm eff} \cdot\left(-\frac{1}{\Gamma_{1}} \frac{\boldsymbol{\nabla} P_{0}}{P_{0}}+\frac{\boldsymbol{\nabla} \rho_{0}}{\rho_{0}}\right).
\end{equation}
Finally, we expand the wave's velocity and fluctuations on Fourier series both in time and in azimuth
\begin{gather}
    \vec{v}(\zeta, \theta, \varphi, t)  \equiv \sum_{\omega^{\mathrm{in}}, m}\left\{\vec{u}(\zeta, \theta) \exp [i(\omega^{\mathrm{in}} t-m \varphi)]\right\},\\
    X^{\prime}(\zeta, \theta, \varphi, t)  \equiv \sum_{\omega^{\mathrm{in}}, m}\left\{\widetilde{X}(\zeta, \theta) \exp [i(\omega^{\mathrm{in}} t-m \varphi)]\right\},
\end{gather}
where $m$ is the azimutal order and $\omega^{\mathrm{in}}$ is the wave frequency in an inertial reference frame. In a differentially  rotating region, the waves are Doppler-shifted due to the differential rotation so we can define the wave frequency $\omega$ in the rotating reference frame as
\begin{equation} \label{eq:doppler_shift}
    \omega(\zeta, \theta) = \omega^{\mathrm{in}} - m\Omega(\zeta, \theta).
\end{equation}

\section{Generalised TAR}\label{sect:Generalised_TAR}
\subsection{Approximation on the stratification profile: $N^2(\zeta, \theta)\approx N^2(\zeta)$} \label{subsect:hypothesis1}
As in Paper I, in order to obtain a separable system of equations when applying the TAR, we assume here that the Brunt-Väisälä frequency $N^2$  depends mainly on the pseudo-radius $\zeta$. This implies also that the hydrostatic pressure $P_0$ and the hydrostatic density $\rho_0$ depend mainly on $\zeta$. We present the validity domain of this approximation in Sect.\;\ref{subsect:validity} using two-dimensional ESTER stellar models.

\subsection{The TAR with centrifugal acceleration in differentially rotation stars}

By assuming the following frequencies hierarchy imposed by the TAR: $\omega\ll N$ and $2\Omega\ll N$ (we verify this hierarchy  in Sect.\;\ref{sect:hierarchy_validation}) which leads to a mostly horizontal velocity equation because of the strong buoyancy restoring force action in the vertical direction ($|u^\zeta|\ll\{|u^\theta|,|u^\varphi|\}$), the approximation $N^2(\zeta, \theta)\approx N^2(\zeta)$, the Cowling approximation ($\Phi^\prime = 0$), and the anelastic approximation (acoustic waves are filtered out), we can rewrite accordingly  the energy equation (Eq.\;\ref{eq:energy_eq}) and the three components of the momentum equation (Eqs.\;\ref{eq:radial_momentum}, \ref{eq:latitudinal_momentum} \& \ref{eq:azimuthal_momentum}).
First, we rewrite the energy equation as
\begin{equation}\label{eq:energy_eq_simplified}
      \frac{\widetilde{\rho}}{\rho_0}=\frac{1}{i\omega}\frac{ N^{2} }{g_{\rm eff}} \frac{\zeta^2}{r^2r_\zeta}u^\zeta,
\end{equation}
where we neglect the term $(1 / \Gamma_{1}) {P}^\prime / P_0$ in Eq.\;(\ref{eq:energy_eq}) using the anelastic approximation and we simplify the expression of the squared Brunt-Väisälä frequency (Eq.\;\ref{eq:N2}) using the approximation \ref{subsect:hypothesis1}. Then, we simplify the radial momentum equation (Eq.\;\ref{eq:radial_momentum}) to
\begin{equation}\label{eq:radial_momentum_TAR}
    i{\omega} \partial_\zeta \widetilde{W}+N^2 \zeta^2 \mathcal{A} u^\zeta=0,
\end{equation}
where we neglect the term $({P}^\prime / \rho_0^{2})  \partial_{\zeta} \rho_0 $ thanks to the anelastic approximation and the vertical component of the acceleration and of the Coriolis acceleration, which are dominated by the buoyancy term since we assume that $\omega \ll N$ and $2\Omega \ll N$ within the TAR. Subsequently, the latitudinal component of the momentum equation (Eq.\;\ref{eq:latitudinal_momentum}) reduces to
    \begin{equation}
    \label{eq:latitudinal_momentum_TAR}
    i{\omega} \zeta \mathcal{B} u^\theta-2 \Omega \zeta \mathcal{C}u^\varphi+\partial_\theta \widetilde{W}=0,
    \end{equation}
where we neglect the terms involving the vertical wave velocity since $|u^\zeta|\ll\{|u^\theta|,|u^\varphi|\}$ because of the strong stable stratification. Finally, the azimuthal component of the momentum equation (Eq.\;\ref{eq:azimuthal_momentum}) simplifies, for the same reasons, into
    \begin{equation}
    \label{eq:azimuthal_momentum_TAR}
    i{\omega} \zeta \mathcal{D} u^\varphi+\left(2\Omega\zeta \mathcal{C} + \frac{\sin{\theta}}{r_\zeta}\partial_\theta \Omega\right)u^\theta-\frac{i m}{\sin{\theta}} \widetilde{W}=0.
    \end{equation} 
The coefficients $\mathcal{A}$, $\mathcal{B}$, $\mathcal{C}$, and $\mathcal{D}$ are defined in Table\,\ref{table:termes} and $\widetilde{W}=\widetilde{P}/\rho_0$ is the normalised pressure.
We thus obtain a system of equations which has the same mathematical form as in the case of: (i) spherically symmetric uniformly rotating \citep{Bildsten1996, lee+saio1997} and differentially rotating \citep{mathis2009} stars, (ii) weakly deformed uniformly rotating \citep{mathis+prat2019} stars and (iii) strongly deformed uniformly rotating (Paper I) stars. We thus still manage to partially decouple the vertical and horizontal components of the velocity. 
By solving the system formed by Eqs.\;(\ref{eq:latitudinal_momentum_TAR}) and (\ref{eq:azimuthal_momentum_TAR}) we can express $u^\theta$ and $u^\varphi$ as a function of $\widetilde{W}$ as follows
\begin{align}\label{eq:u_theta(zeta,x)}
    u^\theta(\zeta, x)&=\mathcal{L}_{\omega^{\mathrm{in}} m}^{\theta}\left[\widetilde{W}(\zeta, x)\right] \nonumber\\
    \begin{split}
    &=-i\frac{1}{\zeta}\frac{1}{\omega(\zeta, x)}\frac{1}{\mathcal{B}(\zeta, x)}\frac{1}{\sqrt{1-x^2}}\\
    &\mathrel{\phantom{w}}\left[\left(1-x^2\right)\left(1+\frac{\nu^2(\zeta, x) \mathcal{C}^2(\zeta, x)/\mathcal{B}(\zeta, x)+\mathcal{F}(\zeta, x)}{\mathcal{E}(\zeta, x)\mathcal{B}(\zeta, x)- \mathcal{F}(\zeta, x)}\right)\partial_x \right. \\
     &\mathrel{\phantom{-i\frac{1}{\omega}\frac{1}{\zeta}\frac{1}{\mathcal{B}(\zeta, x)}\frac{1}{\sqrt{1-x^2}}}} +\left. m\frac{\nu(\zeta, x)  \mathcal{C}(\zeta, x)}{\mathcal{E}(\zeta, x)- \mathcal{F}(\zeta, x)}\right]\widetilde{W},
    \end{split}
\end{align}

\begin{align} \label{eq:u_phi(zeta,x)}
    u^\varphi(\zeta, x) &= \mathcal{L}_{\omega^{\mathrm{in}} m}^{\varphi}\left[\widetilde{W}(\zeta, x)\right] \nonumber\\
    \begin{split}
    &=\frac{1}{\zeta}\frac{1}{\omega(\zeta, x)}\frac{1}{\mathcal{E}(\zeta, x)-\mathcal{F}(\zeta, x)}\frac{1}{\sqrt{1-x^2}}\\
    & \left[\left(1-x^2\right)\left(\frac{ \nu(\zeta, x)\mathcal{C}(\zeta, x)}{\mathcal{B}(\zeta, x)} + \frac{\mathcal{F}(\zeta, x)}{\nu(\zeta, x) \mathcal{C}(\zeta, x)}\right)\partial_x + m \right]\widetilde{W},
    \end{split}
\end{align}
where $\mathcal{E}$ and $\mathcal{F}$ are defined in Table\,\ref{table:termes}, $x=\cos{\theta}$ is the reduced latitudinal coordinate and 
\begin{equation}
    \nu(\zeta, \theta) = \frac{2\Omega(\zeta, \theta)}{\omega(\zeta, \theta)},
\end{equation}
is the spin parameter.

The structure of the new equations that we obtain in the spheroidal differentially rotating case is similar to the one with uniform rotation (Paper I). But we can point out two major differences. First, the wave frequency $\omega$ and the spin parameter $\nu$ are no longer uniform  but they now depend  on the pseudo-radius $\zeta$ and the reduced latitudinal coordinate $x$. That is why it is more convenient to  index the operators and the variables by $\omega^{\mathrm{in}}$ (independent of $\zeta$ and $\theta$) instead of using the spin parameter $\nu$ as in Paper I. Second, a new coefficient $\mathcal{F}$ is involved in the derivation of the dynamics of GIWs that models the action of differential rotation.
\begin{table}
    \centering
    \caption{Terms involved in the derivation of the generalised Laplace tidal equation in the general case of spheroidal geometry and in the particular case of spherical geometry with differential rotation.}
    \label{table:termes}
        \begin{tabular}{c|c|c}
        \hline \hline
        Terms & Spheroidal  & Spherical 
        \\\hline \hline
        $\mathcal{A}$ & $\displaystyle{\frac{1}{r^2 r_\zeta}}$ & $\displaystyle{\frac{1}{\zeta^2}}$ \\\hline
        $\mathcal{B}$ & $ \displaystyle{\frac{r^2+(1-x^2)r_x^2}{r^2r_\zeta}}$  & $1$  \\\hline
        $\mathcal{C}$ & $ \displaystyle{\frac{-(1-x^2) r_x+r x}{r r_\zeta}}$ & $x$  \\\hline
        $\mathcal{D}$ & $\displaystyle{\frac{1}{r_\zeta}}$  & $1$ \\\hline
        $\mathcal{E}$ & $\displaystyle{\mathcal{D}-\frac{\nu^2\mathcal{C}^2}{\mathcal{B}}}$ & $1-\nu^2x^2$   \\\hline
        $\mathcal{F}$ & $\displaystyle{-\frac{\nu }{\omega}\frac{\mathcal{C}\mathcal{D}}{\mathcal{B}}(1-x^2)\partial_x\Omega}$ & $\displaystyle{-\frac{\nu x}{\omega} (1-x^2)\partial_x\Omega}$   \\\hline
        \end{tabular}    
\end{table}

\section{Dynamics of low-frequency gravito-inertial waves}\label{sect:Dynamics_GIWs}
We now derive the generalised Laplace tidal equation for the normalised pressure $\widetilde{W}$, which allows us to compute the frequencies and periods of low-frequency GIWs and to build the corresponding seismic diagnostics. Applying the anelastic approximation and the approximation $N^2(\zeta, \theta)\approx N^2(\zeta)$ in the continuity equation (Eq.\;\ref{eq:continuity}) simplifies it into 
\begin{equation}\label{eq:continuity_simplified}
    \zeta \partial_\zeta \rho_0 u^\zeta+\rho_0\left[\frac{\partial_{\zeta}\left(\zeta^{2} u^{\zeta}\right)}{\zeta}+\frac{1}{\sin{\theta}}\partial_\theta(\sin{\theta}u^\theta)-\frac{i m }{\sin{\theta}}u^\varphi\right]=0.
\end{equation}

\subsection{JWKB approximation}
Under the assumption that $\omega\ll N$, each scalar field and each component of $\vec{u}$ can be expanded using the two-dimensional Jeffreys-Wentzel-Kramers-Brillouin JWKB approximation \citep{froman2005}. In this case, the spatial structure of the waves can be described by the product of a rapidly oscillating plane-like wave function in the pseudo-radial direction multiplied by a slowly varying envelope:
\begin{align}\label{eq:jwkb_w}
\widetilde{W}(\zeta, \theta)&=\sum_{k}\left\{w_{\omega^{\mathrm{in}} k m}(\zeta, \theta) \frac{A_{\omega^{\mathrm{in}} k m}}{k_{V ; \omega^{\mathrm{in}} k m}^{1 / 2}} \exp \left[i \int^{\zeta} k_{V ; \omega^{\mathrm{in}} k m} \mathrm{d} \zeta\right]\right\},
\\u^{j}(\zeta, \theta)&=\sum_{k}\left\{\hat{u}^j_{\omega^{\mathrm{in}} k m}(\zeta, \theta) \frac{A_{\omega^{\mathrm{in}} k m}}{k_{V ; \omega^{\mathrm{in}} k m}^{1 / 2}} \exp \left[i \int^{\zeta} k_{V;\omega^{\mathrm{in}} k m} \mathrm{d} \zeta \right]\right\},\label{eq:jwkb_u}
\end{align}
with $j \equiv\{\zeta, \theta, \varphi\}$, $k$ is the index of a latitudinal eigenmode (cf. Sect.\;\ref{subsect:glte}) and $A_{\omega^{\mathrm{in}} k m}$ is the amplitude of the wave.
Substituting the expansion given in Eqs.\;(\ref{eq:jwkb_w}) and (\ref{eq:jwkb_u}) into Eqs.\;(\ref{eq:radial_momentum_TAR}), (\ref{eq:u_theta(zeta,x)}) and (\ref{eq:u_phi(zeta,x)}), the final pseudo-radial, latitudinal and azimuthal components of the velocity are obtained:
\begin{align}
    \hat{u}^\zeta_{\omega^{\mathrm{in}} k m}(\zeta, x) &=  \frac{ k_{V;\omega^{\mathrm{in}} k m}(\zeta)}{N^2(\zeta)} \frac{\omega_{ k m}(\zeta, x)}{\zeta^2 \mathcal{A}(\zeta, x)} w_{\omega^{\mathrm{in}} k m}(\zeta, x), \label{eq:u_zeta_final}
    \\\hat{u}^\theta_{\omega^{\mathrm{in}} k m}(\zeta, x) &= \mathcal{L}_{\omega^{\mathrm{in}} m}^{\theta}\left[w_{\omega^{\mathrm{in}} k m}(\zeta, x)\right], \label{eq:u_theta_final}
    \\\hat{u}^\varphi_{\omega^{\mathrm{in}} k m}(\zeta, x) &= \mathcal{L}_{\omega^{\mathrm{in}} m}^{\varphi}\left[w_{\omega^{\mathrm{in}} k m}(\zeta, x)\right].\label{eq:u_phi_final}
\end{align}

\subsection{Approximation: $\mathcal{A}(\zeta, \theta)\approx \mathcal{A}(\zeta)$} \label{subsubsect:hypothesis2}
As in Paper I, to be able to introduce the eigenvalues $\Lambda_{\omega^{\mathrm{in}} k m}$ which depends only on $\zeta$ and derive the generalised Laplace tidal equation, we have to do a partial separation between the pseudo-radial and latitudinal variables in the pseudo-radial component of the velocity  (Eq.\;\ref{eq:u_zeta_final}). So, we have to assume that the coefficient $\mathcal{A}$ depends mainly on the pseudo-radius $\zeta$. The validity domain of this approximation is presented in Sect.\;\ref{subsect:validity} using two-dimensional ESTER stellar models. Unlike the uniformly rotating case (Paper I), this assumption alone is insufficient to partially separate the variables in Eq.\;(\ref{eq:u_zeta_final}) in the differentially rotating case), so it is mandatory to assume a third approximation. 

\subsection{Approximation: $\Omega(\zeta, \theta)\approx \Omega(\zeta)$} \label{subsubsect:hypothesis3}
The wave frequency in the rotating frame  (Eq.\;\ref{eq:doppler_shift}) depends on $\zeta$ and $\theta$. This dependency comes from the differential rotation $\Omega(\zeta,\theta)$. Therefore, in order to preform the partial  separation of variables in Eq.\;(\ref{eq:u_zeta_final}), the angular velocity $\Omega$ should have a dependency only on $\zeta$. We examine in detail the validity of this approximation for two-dimensional ESTER stellar models in Sect.\,\ref{subsect:validity_hypo}.

\subsection{Generalised Laplace tidal equation} \label{subsect:glte}
Substituting the expansion given in Eqs.\;(\ref{eq:jwkb_w}) and (\ref{eq:jwkb_u}) into Eqs.\;(\ref{eq:radial_momentum_TAR}), (\ref{eq:u_theta(zeta,x)}) and (\ref{eq:u_phi(zeta,x)}), then replacing them into the continuity equation (Eq.\;\ref{eq:continuity_simplified}) we obtain the generalised Laplace tidal operator (GLTO)
\begin{multline}
    \mathcal{L}_{\omega^{\mathrm{in}} m} = \omega \partial_x\left[\frac{1}{\omega}\frac{1}{\mathcal{B}(\mathcal{E}-\mathcal{F})}\left(\mathcal{E}+\frac{\nu^2\mathcal{C}^2}{\mathcal{B}}\right)(1-x^2)\partial_x \right]
    -\\\frac{m\mathcal{F}}{\nu \mathcal{C}(\mathcal{E}-\mathcal{F})}\partial_x 
    +m \omega \partial_x\left(\frac{\nu \mathcal{C}}{\omega\mathcal{B}(\mathcal{E}-\mathcal{F})}\right) -\frac{m^2}{(\mathcal{E}-\mathcal{F})(1-x^2)},
    \label{eq:glto}
\end{multline}
thus the generalised Laplace tidal equation (GLTE) for the normalised pressure $w_{\omega^{\mathrm{in}} k m}$ is written as
\begin{equation}
    \mathcal{L}_{\omega^{\mathrm{in}} m}\left[w_{\omega^{\mathrm{in}} k m}\right]= -\Lambda_{\omega^{\mathrm{in}} k m}(\zeta) w_{\omega^{\mathrm{in}} k m},\label{eq:glte}
\end{equation}
where $\Lambda_{\omega^{\mathrm{in}} k m}$ are the eigenvalues deduced from the following dispersion relation:
\begin{equation}\label{eq:dispersion}
    k_{V ; \omega^{\mathrm{in}} k m}^{2}(\zeta)=\frac{N^2(\zeta)\mathcal{A}(\zeta)}{\omega_{k m}^{2}(\zeta)} \Lambda_{\omega^{\mathrm{in}} k m}(\zeta),
\end{equation}
and $w_{\omega^{\mathrm{in}} k m}$ the generalised Hough functions (eigenfunctions)  of the GLTE. We choose to define our latitudinal ordering number $k$ to enumerate, for each $(\omega^{\mathrm{in}},m)$, the infinite set of solutions as in \citet{lee+saio1997}, \citet{mathis2009} and \citet{mathis+prat2019}.

The GLTO (Eq.\;\ref{eq:glto}) reduces to the Laplace
tidal operator in the case of a uniformly rotating deformed star (Paper I). In this case, $\omega$ and $\nu$ are constants so we can take them out of the derivatives and $\mathcal{F}=0$; thus we obtain 
\begin{multline}
    \mathcal{L}_{\nu m} = \partial_x\left[\frac{1}{\mathcal{B}}\left(1+\frac{\nu^2\mathcal{C}^2}{\mathcal{E}\mathcal{B}}\right)(1-x^2)\partial_x \right] +\\\left(m \nu \partial_x\left(\frac{ \mathcal{C}}{\mathcal{E}\mathcal{B}}\right) -\frac{m^2}{\mathcal{E}(1-x^2)}\right).
\end{multline}
Furthermore, the GLTO (Eq.\;\ref{eq:glto}) reduces to the Laplace tidal operator derived by \cite{mathis2009} in the case of a spherical differentially rotating star. In this case, we can replace coefficients  $\mathcal{A}$, $\mathcal{B}$, $\mathcal{C}$, $\mathcal{D}$, and $\mathcal{E}$ by their analytical expressions in the particular case of a spherical geometry presented in Table\,1 of Paper I, so the new coefficient $\mathcal{F}$ defined here can be written in this particular case as 
\begin{equation}
    \mathcal{F}(\zeta,x)= -\frac{\nu x}{\omega} (1-x^2)\partial_x\Omega,
\end{equation}
thus we obtain 
\begin{multline}
    \mathcal{L}_{\omega^{\mathrm{in}} m} = \omega \partial_x\left[\frac{1-x^2}{\omega\mathcal{G}}\partial_x \right] - \frac{m\partial_x\Omega}{\omega\mathcal{G}}(1-x^2)\partial_x +
    \\ \left(m \omega \partial_x\left(\frac{\nu x}{\omega\mathcal{G}}\right) -\frac{m^2}{\mathcal{G}(1-x^2)}\right),
\end{multline}
with 
\begin{equation}
    \mathcal{G}(\zeta,x)= \mathcal{E}(\zeta,x)-\mathcal{F}(\zeta,x)= 1-\nu^2x^2+\frac{\nu x}{\omega} (1-x^2)\partial_x\Omega.
\end{equation}

\subsection{Asymptotic frequency and period spacing of low-frequency GIWs}
By substituting the dispersion relation (Eq.\;\ref{eq:dispersion}) into the following  quantisation condition in the vertical direction defined by \cite{unno1989}, \cite{gough1993}, and \cite{Christensen1997} and used in Paper I
\begin{equation}\label{eq:quantisation}
    \int_{\zeta_{1}}^{\zeta_{2}} k_{V ; \omega^{\mathrm{in}} n k m} \mathrm{d} \zeta=(n+1 / 2) \pi,
\end{equation}
where $\zeta_1$ and $\zeta_2$ are the turning points of the 
Brunt–Väisälä frequency $N$ and $n$ is the radial order, we compute numerically the asymptotic frequencies of low-frequency GIWs in deformed
differentially rotating stars in the inertial frame. We describe in detail the used numerical method in Sec.\;\ref{sect:seismic_diagnosis}. In the case of a weak differential rotation, we can extract, analytically, the expression of the asymptotic frequencies in the rotating frame
\begin{equation}\label{eq:frequencies}
    \omega_{n k m}=\frac{\int_{\zeta_{1}}^{\zeta_{2}} N(\zeta)\sqrt{\mathcal{A}(\zeta)\Lambda_{\omega^{\mathrm{in}}_n k m}(\zeta)} \mathrm{d} \zeta}{(n+1 / 2) \pi},
\end{equation}
with 
\begin{equation}
    \omega^{\mathrm{in}}_n= \omega^{\mathrm{in}}_{n k m}= \omega_{n k m} + m\Omega_{\rm av},
\end{equation}
where $\Omega_{\rm av}$ is here the averaged rotation. Then by applying a first-order Taylor development following \cite{Bouabid2013}, we can generalise his expression of the corresponding period spacing obtained in the spherical case
\begin{align}\label{eq:period_spacing}
    \Delta P _{km}&= P_{nkm}- P_{n-1km} \approx\displaystyle{\frac{2 \pi^2}{\int_{\zeta_{1}}^{\zeta_{2}} N\sqrt{\mathcal{A}\Lambda_{\omega^{\mathrm{in}}_{n+1} k m}} \mathrm{d} \zeta \left(1+\alpha\right)}},
\end{align}
 with 
 \begin{equation}
     \alpha= \frac{1}{2}\frac{\int_{\zeta_{1}}^{\zeta_{2}} N\sqrt{\mathcal{A}\Lambda_{\omega^{\mathrm{in}}_{n} k m}} \frac{d \ln \Lambda_{\omega^{\mathrm{in}}_{n} k m}}{d \ln{\omega^{\mathrm{in}}}} \mathrm{d} \zeta}{\int_{\zeta_{1}}^{\zeta_{2}} N\sqrt{\mathcal{A}\Lambda_{\omega^{\mathrm{in}}_{n} k m}} \mathrm{d} \zeta}.
 \end{equation}
\section{Application to rapidly and differentially rotating early-type stars}\label{sect:results}
As in Paper I, we use here also ESTER models \citep{espinosa+rieutord2013} to implement our equations. We use $3 M_{\odot}$ stellar models with a hydrogen mass fraction in the core $X_{c} = 0.7$ rotating at the equator at $[0\%,90\%]$ of the Keplerian break-up rotation rate. 

\subsection{Domain of validity of the generalised TAR} \label{subsect:validity}
We study, within the framework of ESTER models, the validity domain of the three approximations  (\ref{subsect:hypothesis1}, \ref{subsubsect:hypothesis2} and \ref{subsubsect:hypothesis3}) that are necessary to build the TAR in the case of rapidly differentially  rotating deformed bodies:
\begin{gather}
    N^2(\zeta, \theta)\approx N^2  (\zeta),\label{eq:approx1}\\
    \mathcal{A}(\zeta, \theta)\approx\mathcal{A}(\zeta),\label{eq:approx2}\\
    \Omega(\zeta, \theta)\approx \Omega(\zeta)\label{eq:approx3}.
\end{gather}
The validity of the  approximations (\ref{eq:approx1}) and (\ref{eq:approx2}) is well discussed in Paper I (the $N^2$ and $\mathcal{A}$ profiles are the same in the uniformly and differentially rotating cases). Here, we focus on determining the validity domain of the approximation (\ref{eq:approx3}) by applying the same method used to evaluate the other two assumptions that we recall here.

\subsubsection{Validity of $\Omega(\zeta, \theta)\approx \Omega(\zeta)$} \label{subsect:validity_hypo}
\begin{figure}
    \centering
    \resizebox{\hsize}{!}{\includegraphics{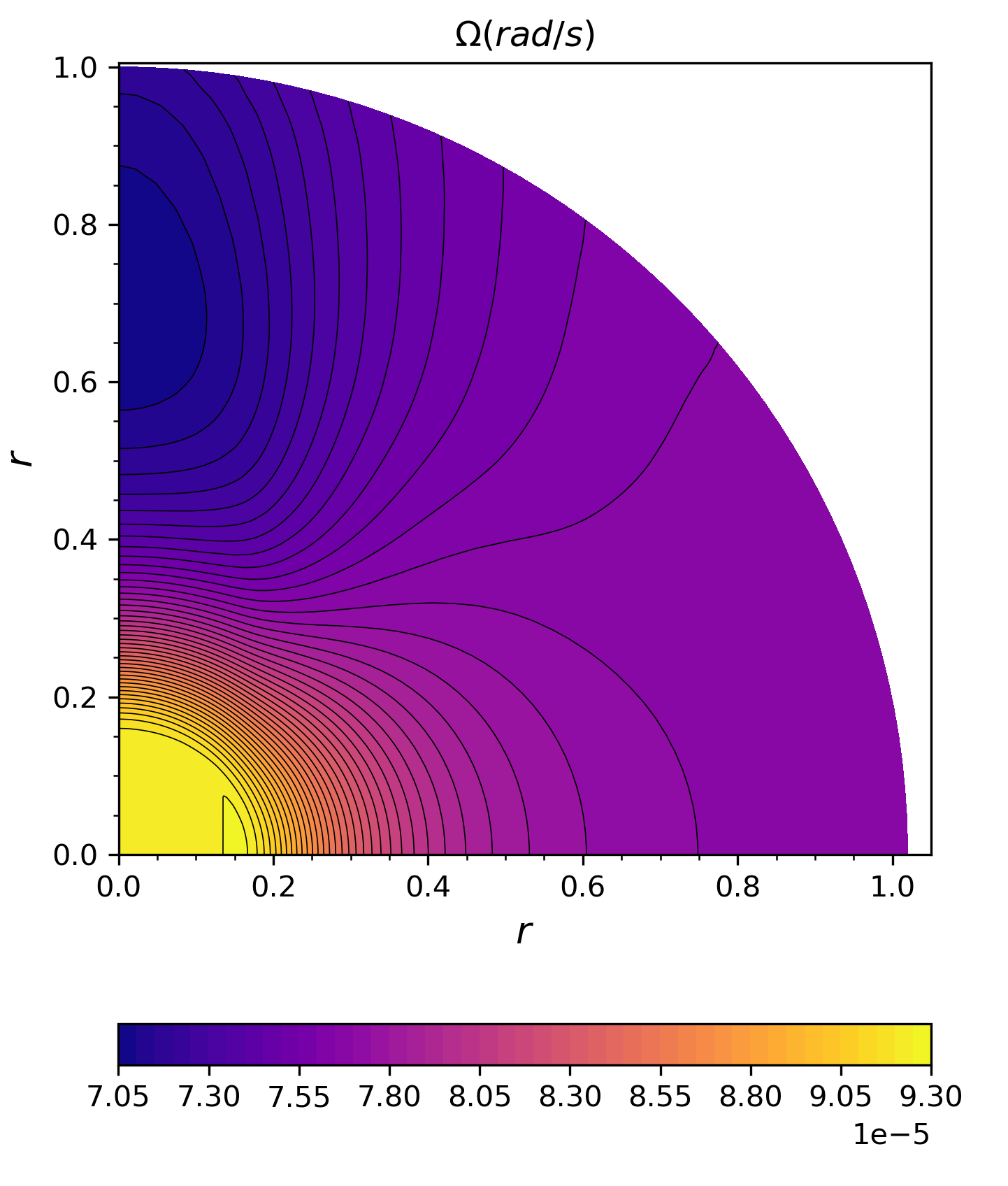}}
    \caption{Angular velocity profile $\Omega(\zeta,\theta)$ 
    of an ESTER model ($3\,\mathrm{M}_{\odot}$, $X_{\mathrm{c}}=0.7$) rotating at the equator at $\Omega/\Omega_{\rm K}=20\%$ (the isolines are represented in black).}
    \label{fig:omega_profile}
\end{figure}

\begin{figure}
    \centering
    \resizebox{\hsize}{!}{\includegraphics{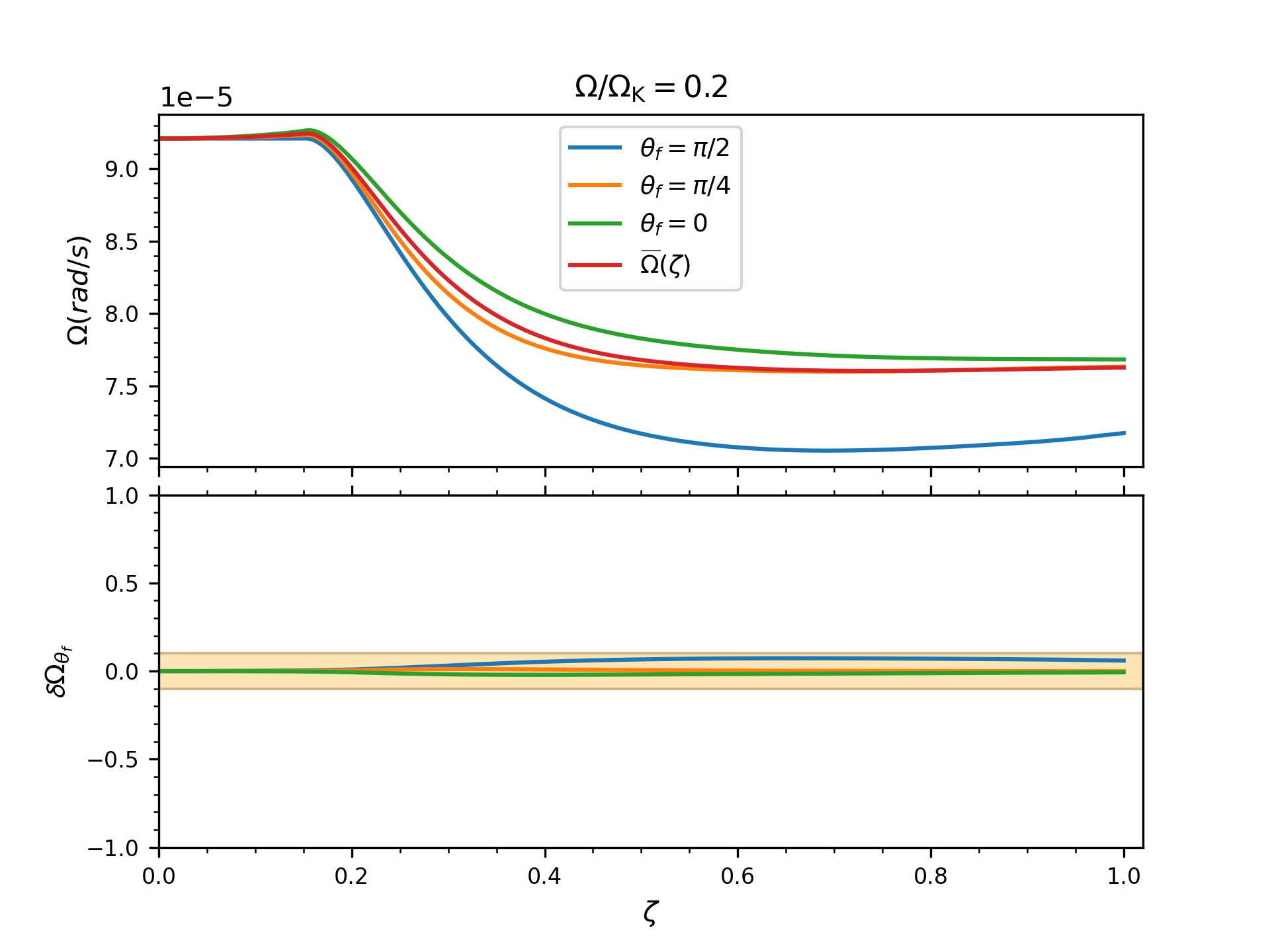}}
    \resizebox{\hsize}{!}{\includegraphics{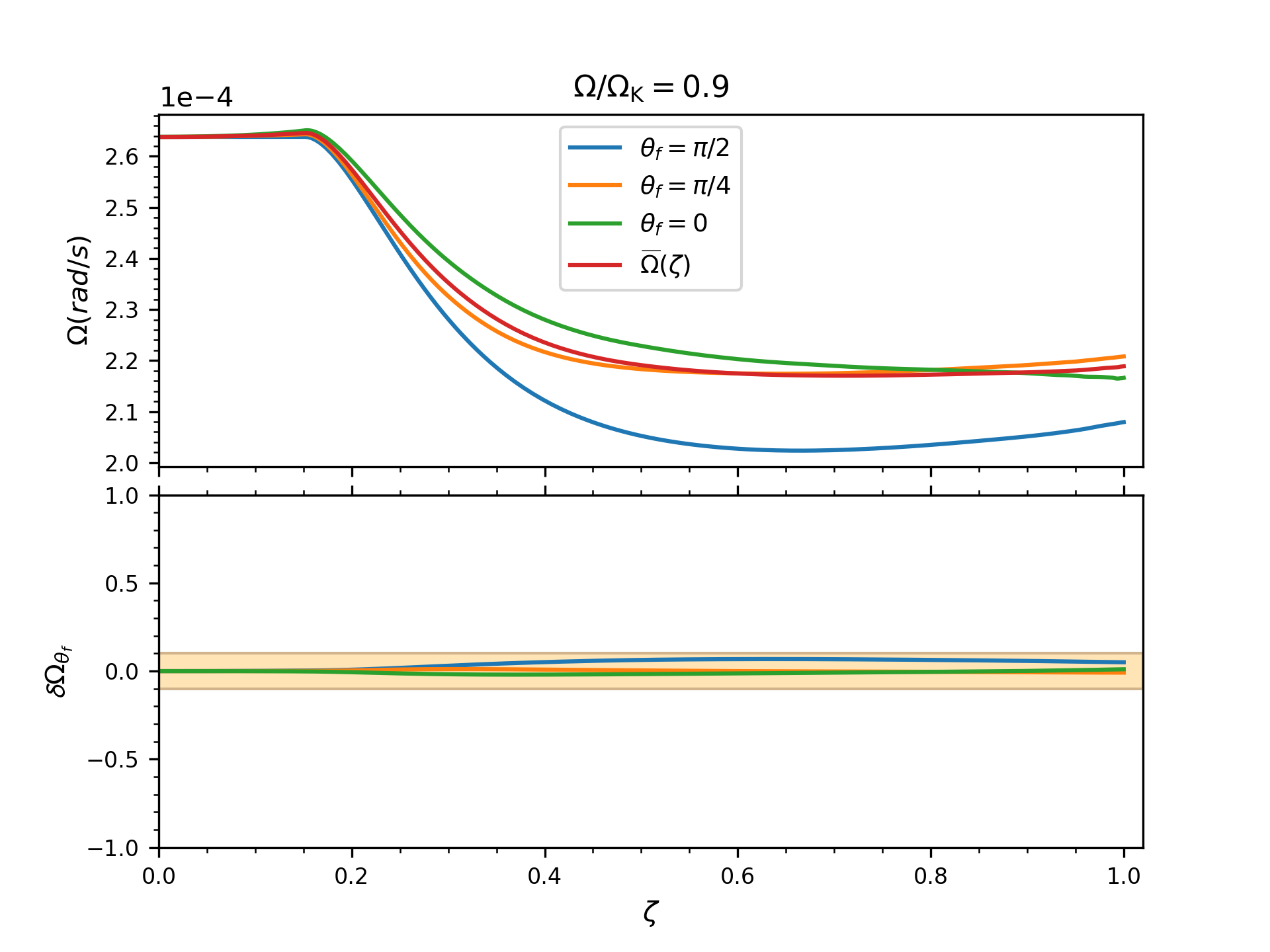}}
    \caption{Profile of the angular velocity $\Omega$ and the relative error $\delta \Omega_{\theta_f}$ of the approximation $\Omega(\zeta, \theta)\approx \Omega(\zeta)$ as a function of $\zeta$ at different colatitudes $\theta_f$ using an ESTER model rotating at the equator at $20\%$ (above) and $90\%$ (below) of the Keplerian break-up rotation rate (the light orange area indicate the margin of error which we allow).}
    \label{fig:omega_approx}
\end{figure}

To visualise the problem, first we illustrate in Fig.\;\ref{fig:omega_profile} the function $\Omega(\zeta, \theta)$ computed with an ESTER model ($3\,\mathrm{M}_{\odot}$, $X_{\mathrm{c}}=0.7$) rotating at the equator at $\Omega/\Omega_{\rm K}=20\%$. We can see that the angular velocity is equal to  $\unit{14.66}{\micro\hertz} $ at the core, and at the surface we get  $\unit{11.41}{\micro\hertz}$  at the pole and $\unit{12.22}{\micro\hertz}$ at the equator. Then, to evaluate the committed error by adopting the approximation
\begin{equation}
    \Omega(\zeta, \theta)\approx \Omega(\zeta),
\end{equation}
we represent in Fig.\;\ref{fig:omega_approx} the angular velocity profile as a function of the pseudo-radius $\zeta$ for different values of the colatitude $\theta_f$, at a fixed rotation rate at the equator $\Omega=0.2 \Omega_{\rm K}$. Visually,  we can notice that the latitudinal gradient of the angular velocity is less important than the radial gradient. In order to confirm this observation, we show, in the bottom panels, the  corresponding relative error $\Delta \Omega_{\theta_f} (\zeta)$ between the exact value $\Omega(\zeta,\theta_f)$ given by the model and the approximated model $\Omega_{\rm approx}(\zeta)$ (the weighted average of the exact value over the colatitude $\theta$, $\bar{\Omega}(\zeta)$). We define these quantities as follows:
\begin{equation}
    \delta \Omega_{\theta_f} (\zeta)= \frac{\Omega_{\rm approx}(\zeta) - \Omega(\zeta,\theta_f)}{\Omega(\zeta,\theta_f)},
\end{equation}
with
\begin{align}
    \Omega_{\rm approx}(\zeta)=\bar{\Omega}(\zeta)&= \frac{\int_0^{\pi/2}\Omega(\zeta,\theta)\sin\theta d\theta}{\int_0^{\pi/2}\sin\theta d\theta}\nonumber\\ &=\int_0^{\pi/2}\Omega(\zeta,\theta)\sin\theta d\theta,
\end{align}
where  $\theta_f$ is a fixed value of the colatitude $\theta$ and we used the symmetry propriety of $\Omega$  with respect to the equator. We can conclude that this approximation is valid with a level of confidence of $90\%$ for all pseudo-radii $\zeta$ and for all rotation rates $\Omega/\Omega_{\rm K}\in \left]0\%, 90\%\right]$ since the error rate is always lower than the maximal error rate fixed to $10\%$. We also recall here that within the physics treated in the ESTER code the differential rotation results from the baroclinic meridional circulation \citep{zahn1992, Maeder+zahn1998, mathis+zahn2004, Rieutord2006} and an isotropic viscous transport. A weaker dependence of $\Omega$ on $\theta$ can be enforced by horizontal turbulent transport \citep{zahn1992, Maeder2003, Mathis2004, mathis2018, park2021} or magnetic fields \citep{Mestel+Weiss1987}.

\subsubsection{Validity domain of the TAR}
\begin{figure}
    \centering
    \resizebox{\hsize}{!}{\includegraphics{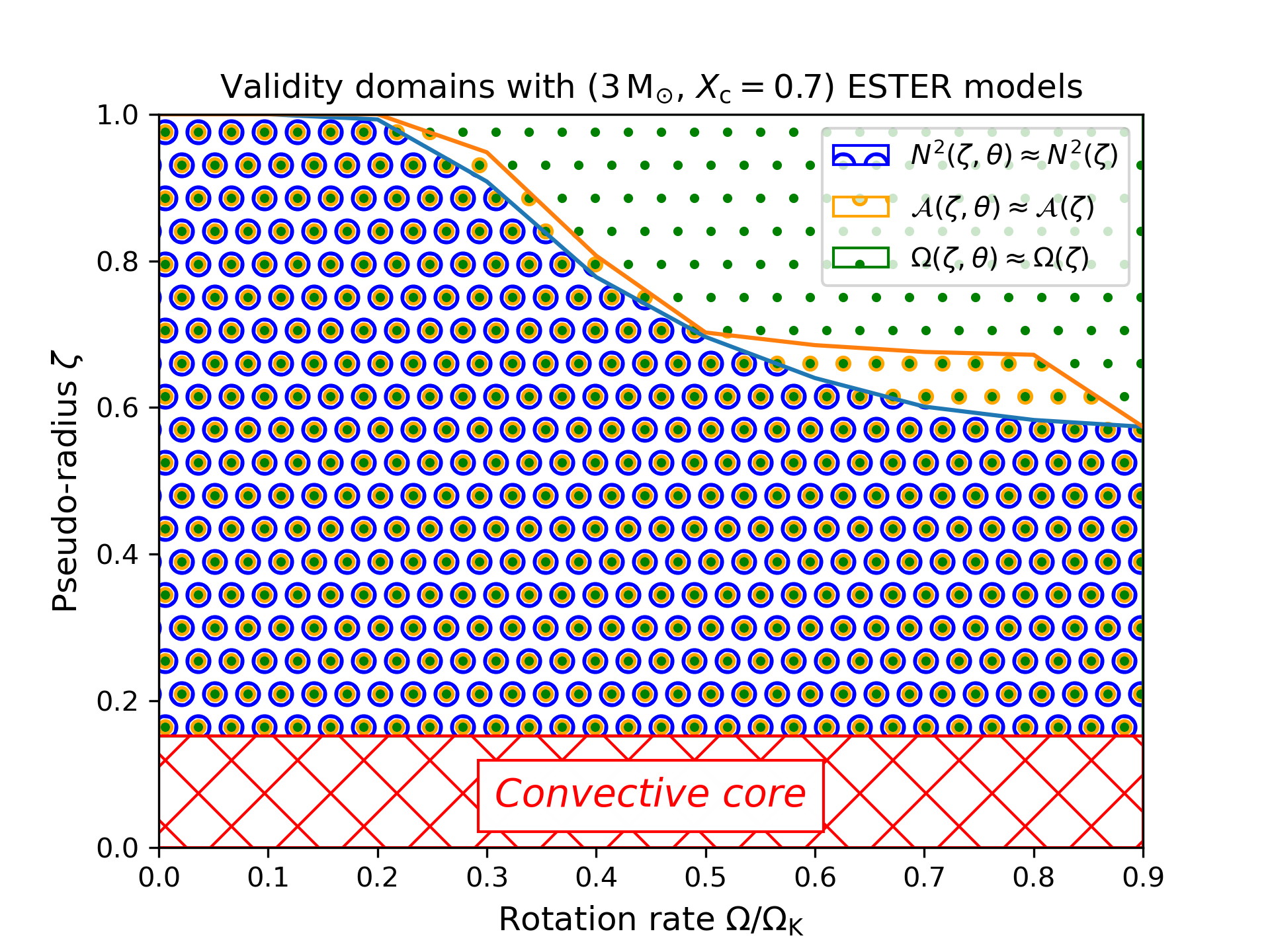}}
    \resizebox{\hsize}{!}{\includegraphics{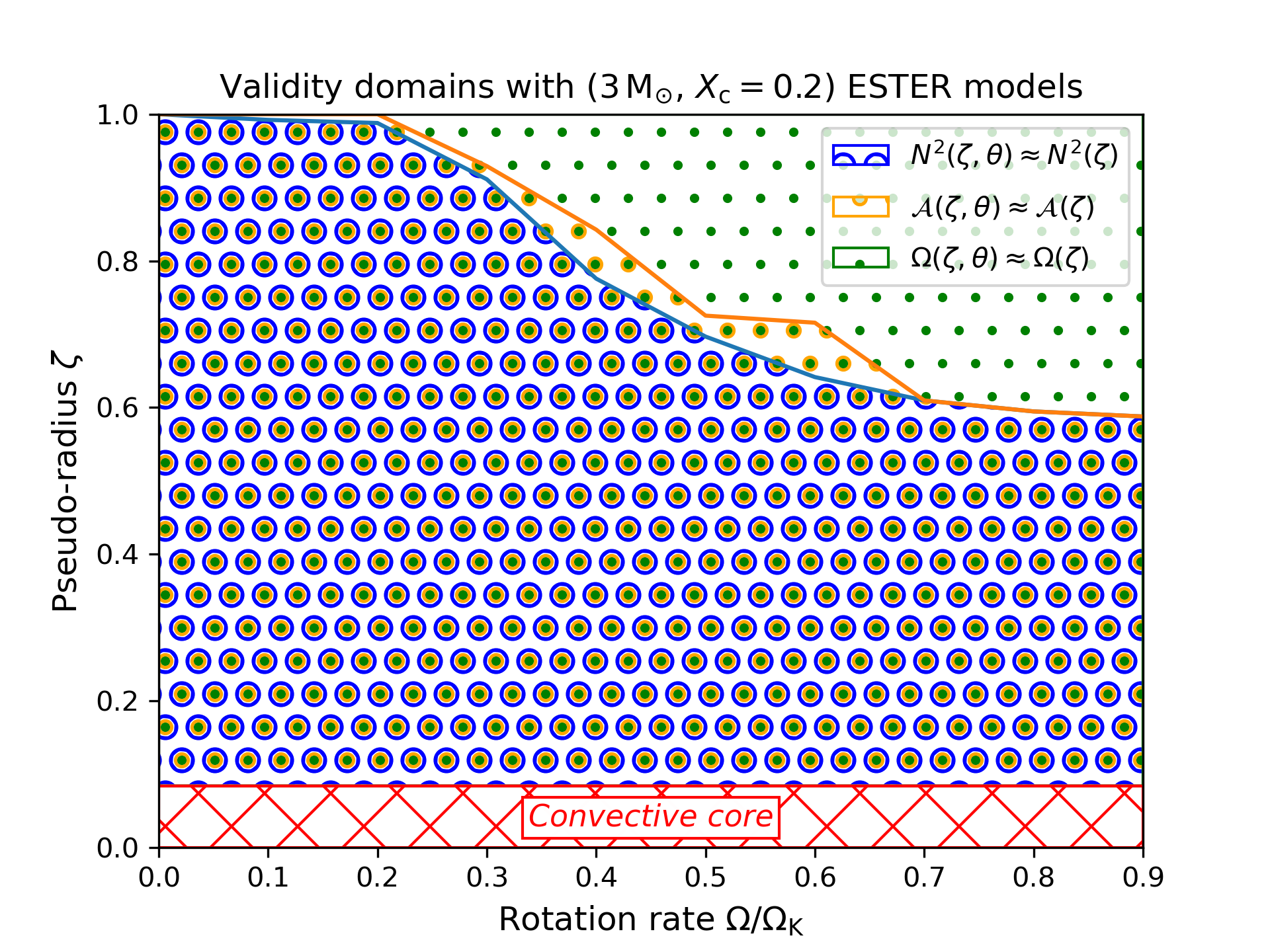}}
    \caption{Validity domain of the TAR within the framework of $3\,\mathrm{M}_{\odot}$, $X_{\mathrm{c}}=0.7$ (above) and $3\,\mathrm{M}_{\odot}$, $X_{\mathrm{c}}=0.2$ (below) ESTER models  as a function of the rotation rate $\Omega /\Omega_{\rm K}$ and the pseudo-radius $\zeta$ (with $90\%$ degree of confidence).}
    \label{fig:domain}
\end{figure}

We determine the validity domain of the three approximations (\ref{eq:approx1}), (\ref{eq:approx2}), and (\ref{eq:approx3}) as a function of the rotation rate and the pseudo-radius by varying systematically the normalised rotation rate of the star $\Omega/\Omega_{\rm K}$  and by calculating for each case the associated maximum relative error. Afterwards, we fix a maximum error rate equal to $10\%$ and we deduce the pseudo-radius limit $ \zeta_{\rm limit} $ where the maximum relative error exceeds this threshold and we decide that the approximations become invalid. Physically, it means that the variation of the structural properties and rotation profile are weak in the latitudinal direction and vary mostly with the pseudo-radius that allows us to build the TAR. The best method in the future to improve the control of the adopted assumptions would be to compute asymptotic frequencies using the latitudinally averaged quantities within the TAR and to compare them to frequencies computed using 2D oscillation codes applied to deformed models \citep[e.g.][]{Ouazzani2015,ouazzani2017,Reese2021}.
In Fig.\;\ref{fig:domain}, we display the pseudo-radius limit $ \zeta_{\rm limit} $ as a function of the rotation rate $\Omega/\Omega_{\rm K}$ for 3 solar masses ESTER models with $X_c=0.7$ and $X_c=0.2$. This curve defines the validity domains of the approximations (\ref{eq:approx1}), (\ref{eq:approx2}), and (\ref{eq:approx3}) which consequently define the validity domain of the TAR.  We note that the influence of the hydrogen mass fraction in the core $X_c$ on the validity domain of the TAR is very weak. It is therefore possible to apply the generalised TAR to differentially rotating early-type stars rotating up to 20\% of the Keplerian critical angular velocity. This limit is the same as the one found in Paper I, so the differential rotation have no influence on the validity domain of the TAR. Chemical gradients created near the core along the evolution are not treated in 2D ESTER models yet. Their treatment can be foreseen by adding them 'by hand' to explore their effects as this has been done for instance for acoustic glitches by \cite{Reese2014,Reese2021}.

\subsection{Eigenvalues and Hough functions}
We solve the GLTE for different pseudo-radii, wave frequency in the inertial frame, and rotation rates within the scope of the defined validity domain using an implementation based on Chebyshev polynomials \citep{wang2016} as in Paper I. 
Here, we are no longer able to calculate the spectrum of the GLTE as a function of the spin parameter $\nu$ because in the differentially rotating case, $\nu$ is no longer a constant value but it becomes a function of $\zeta$ and $\theta$. So to make the comparison between the new results and our previous results clear, we define the weighted average of the spin parameter
\begin{equation}
    \bar{\nu}=\int_0^1\int_0^{\pi/2}\nu(\zeta,\theta)\sin\theta d\theta d\zeta.
\end{equation}
\begin{figure}
    \centering
     \resizebox{\hsize}{!}{\includegraphics{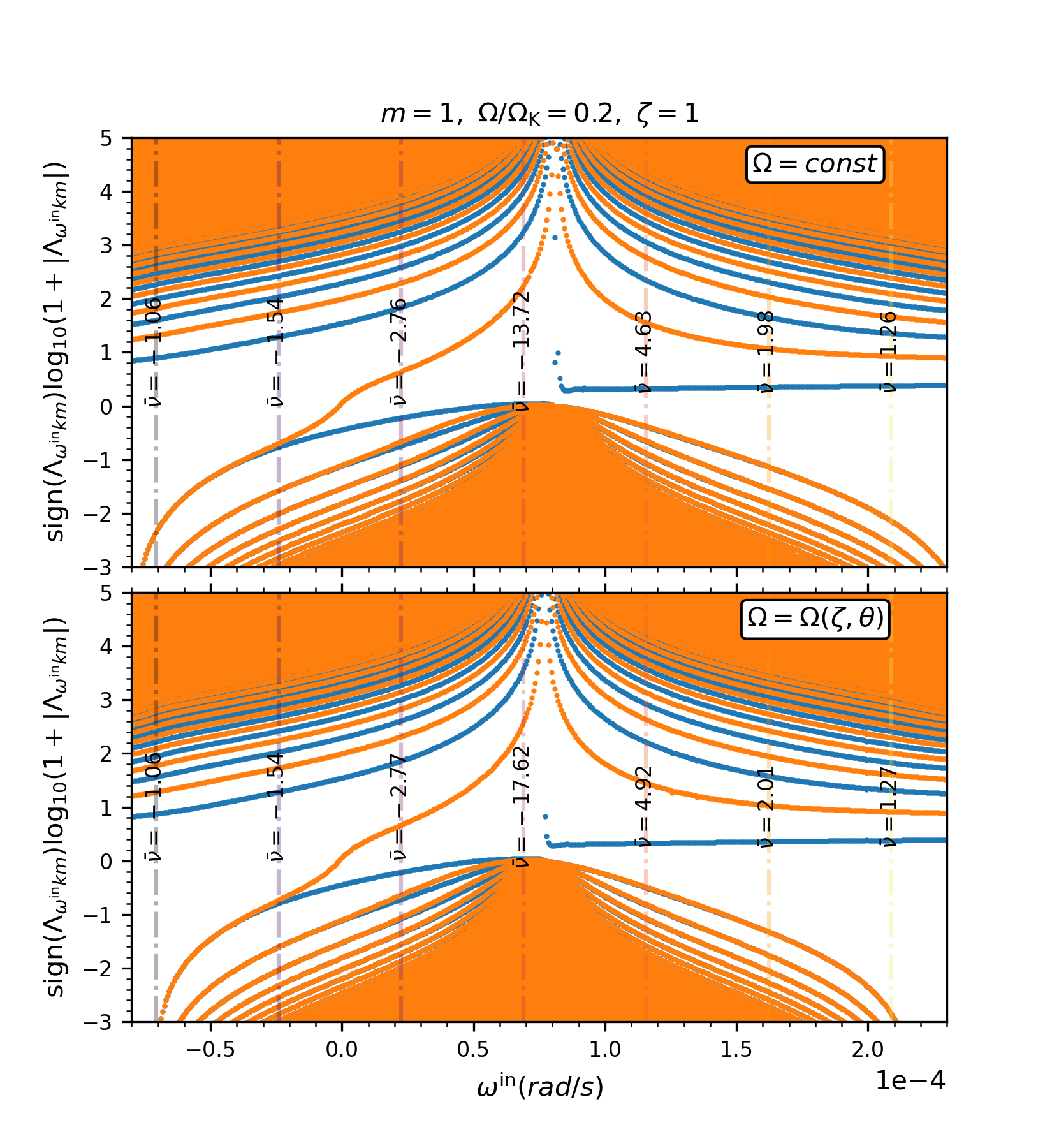}}
    \caption{Spectrum of the GLTE as a function of the wave frequency in the inertial frame $\omega^{\rm in}$ at $\zeta=1$ and $m=1$ for $\Omega=0.2 \Omega_{\mathrm{K}}$ in the case of a  uniformly rotating star (above) and of a differentially rotating star (below).  Blue (respectively, orange) dots correspond to even (respectively, odd) eigenfunctions. The vertical lines correspond to the mean values of the spin parameter $\bar{\nu}$.}
    \label{fig:spectrum_z1_o0.2}
\end{figure}
Fig.\;\ref{fig:spectrum_z1_o0.2} shows the eigenvalues $ \Lambda_{\omega^{\rm in} km}$ as a function of $\omega^{\rm in}$ for $m = 1$ and $\zeta=\zeta_{\rm limit}=1$ at $\Omega/\Omega_{\mathrm{K}}=0.2 $.  Since $\vec{\Omega}$ points in the direction of $\theta =0$ and the oscillations are proportional to $e ^{i (\omega t - m \varphi)}$, a prograde (retrograde) oscillation correspond to  a positive (negative) value of the product $m\nu$ which depends on the spatial coordinates. This figure reveals that the differential rotation of the star causes at $\zeta=1$ a modest gradual shift in the eigenvalues.
We recall here the two main families of eigenvalues. The first one being gravity-like solutions ($\Lambda_{\omega^{\mathrm{in}} k m}\geqslant0$). They exist for any value of $\omega^{\rm in}$ and we attach positive $k$'s to them. They correspond to gravity waves ($\mathrm{g}$ modes) modified by rotation. The second one is Rossby-like solutions. They exist only for specific values of wave frequency in the inertial reference frame ($\omega^{\rm in}\in [\omega^{\rm in}_-, \omega^{\rm in}_+]$) and we attach negative $k$'s to them. They appear only in rotating stars and they correspond to Rossby modes ($\mathrm{r}$ modes) if they are retrograde and have positive eigenvalues ($\Lambda_{\omega^{\mathrm{in}} k m}>0$) and to overstable convective modes if they are prograde and have negative eigenvalues ($\Lambda_{\omega^{\mathrm{in}} k m}<0$).

\begin{figure}
    \centering
     \resizebox{\hsize}{!}{\includegraphics{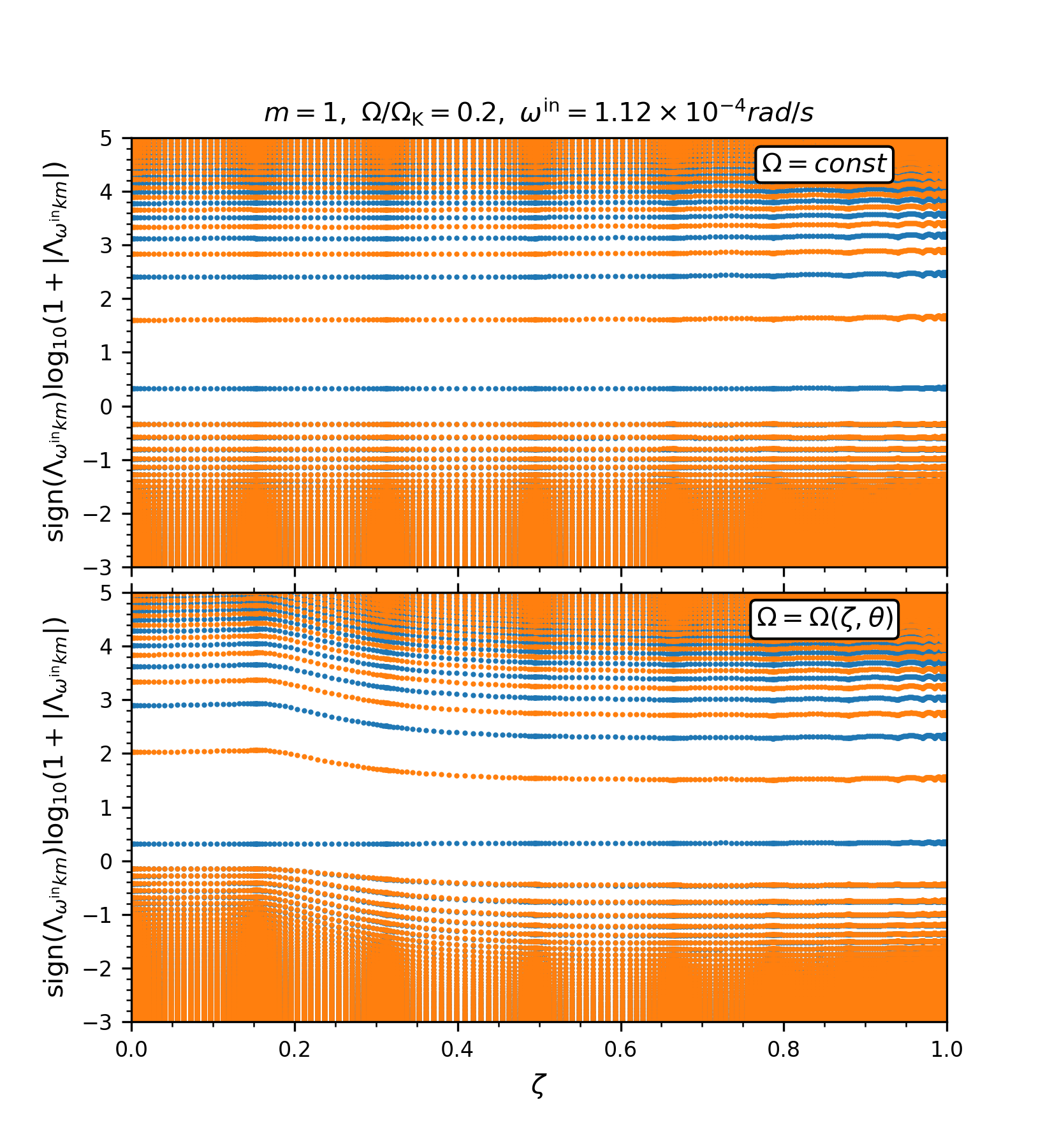}}
    \caption{Spectrum of the GLTE as a function of the pseudo-radius for $m=1$ and $\omega^{\rm in}=1.12\times 10^{-4}~\radian\per\second$ at $ \Omega/\Omega_K = 0.2 $ in the case of a uniformly rotating star (above) and of a differentially rotating star (below). Blue (respectively, orange) dots correspond to even (respectively, odd) Hough functions.}
    \label{fig:spectrumz_nu5}
\end{figure}

Contrary to the uniformly rotating case we obtain here eigenvalues which depend considerably on the pseudo-radius $\zeta$ as shown in Fig.\;\ref{fig:spectrumz_nu5} where we represent the spectrum of the GLTE as a function of the pseudo-radius at $\Omega/\Omega_{\mathrm{K}}=0.2 $, $m=1$ and $\omega^{\rm in}=1.12\times 10^{-4}~\radian\per\second$.
More precisely we can see that the differential rotation influence especially the region close to the convective core ($0.153\le \zeta \le 0.35$) where the rotation is maximal and the radial gradient of the angular velocity is the highest (cf. Fig.\;\ref{fig:omega_profile}). We can see that this dependency becomes very low in the external region $0.5\le \zeta \le 1$ where we obtain eigenvalues very close to the ones in the uniformly rotating case. 

\begin{figure*}
    \centering
     \resizebox{\hsize}{!}{\includegraphics{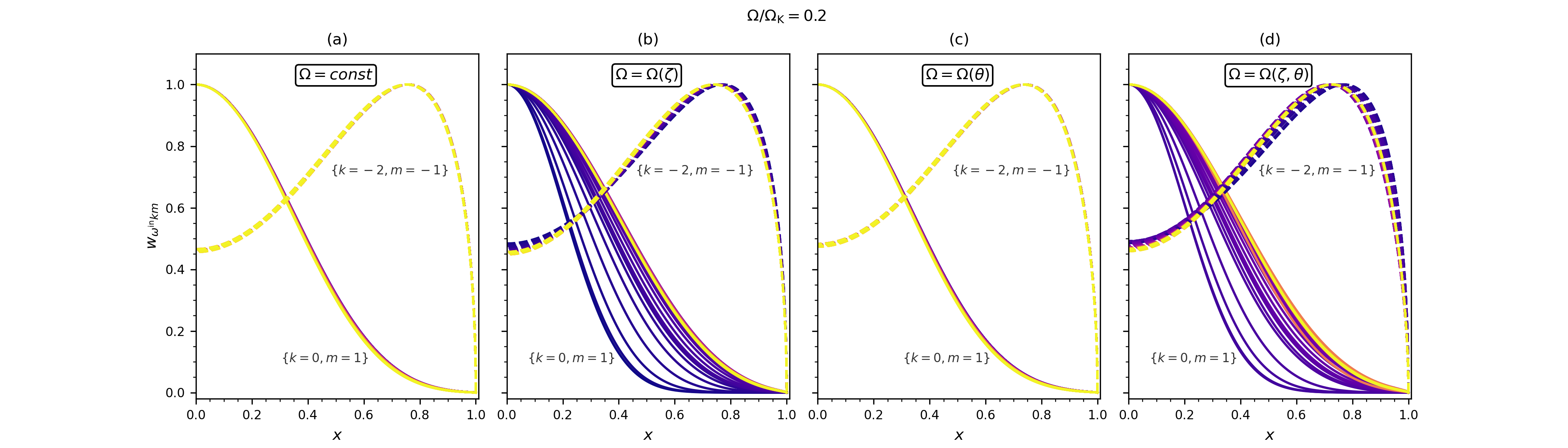}}
    \caption{Generalised Hough functions (normalised) as a function of the horizontal coordinate $x$ at different pseudo-radii from $\zeta=0.153$ (blue) to $\zeta=1 $ (yellow) at $ \Omega/\Omega_K = 0.2 $ for gravity-like solutions $\{k=0, m = 1\} $ with $\omega^{\rm in}= 10^{-4}~\radian\per\second$ corresponding to $\bar{\nu}=10.62$ (the solid lines) and Rossby-like solutions $\{k=-2, m =-1\}$ with $\omega^{\rm in}= -5.5\times 10^{-5}~\radian\per\second$ corresponding to $\bar{\nu}=6.48$ (the dotted lines) in the case of a star: (a) uniformly rotating, (b) differentially rotating according to the pseudo-radius, (c) differentially rotating according to the colatitude and (d) differentially rotating according to the pseudo-radius and the colatitude.}
    \label{fig:hough_nu5}
\end{figure*}

We focus now on the influence of the differential rotation on the generalised Hough functions $w_{km}$ which varies with the pseudo-radius $\zeta$ and the horizontal coordinate $x$. This dependence is illustrated in Fig.\;\ref{fig:hough_nu5} at $ \Omega/\Omega_K = 20\% $ for prograde dipole $\{k=0, m = 1\}$ modes with $\omega^{\rm in}= 10^{-4}~\radian\per\second$ and retrograde  Rossby $\{k=-2, m =-1\}$ modes with $\omega^{\rm in}= -5.5\times 10^{-5}~\radian\per\second$ for four different cases: Fig.\;\ref{fig:hough_nu5}a represents our starting point where the rotation is uniform (Paper I); in Fig.\;\ref{fig:hough_nu5}b, we take into account the pseudo-radial differential rotation; in Fig.\;\ref{fig:hough_nu5}c, we take into account the latitudinal differential rotation and Fig.\;\ref{fig:hough_nu5}d represents our final point where we take into account the full differential rotation profile. As the pseudo-radius decreases from the surface ($\zeta=1$) to the edge of the radiative zone ($\zeta=0.153$), Rossby-like solutions are slightly modified, whereas gravity-like solutions considerably change. We can also see that the major modification comes from the differential rotation gradient along the pseudo-radius $\zeta$ (Fig.\;\ref{fig:hough_nu5}b) while the differential rotation according to the colatitude $\theta$ modifies marginally the gravity-like solutions and has almost no impact on the Rossby-like solution (Fig.\;\ref{fig:hough_nu5}c). On top of that, we can see that the Hough functions at the surface are not too affected  in all the cases and that the strongest variation occur in the interior of the star where the pseudo-radial gradient of the rotation is high.

Overall, we can see clearly in Fig.\;\ref{fig:hough_nu5} that the differential rotation strongly influences and modifies the eigenfunctions of the GLTE by introducing a new dependency on the pseudo-radial coordinate $\zeta$, whereas in the spherically symmetric uniformly rotating case they were only dependant on the latitudinal coordinate $\theta$ while they were also slightly dependent on $\zeta$ in the uniformly rotating deformed case. In fact, the gravity-like and the Rossby-like solutions shift as the distance from the border of the radiative zone of the star ($\zeta=0.153$) to its surface ($\zeta=1$) increases. More specifically, we observe that the gravity-like solutions migrate onwards, away from the equator ($x = 0$), causing a broadening of its shape. This corresponds to the modification of the region of propagation of gravito-inertial waves by the differential rotation \citep{mathis2009, Mirouh2016, Prat2018}. Indeed, since the angular velocity $\overline\Omega(\zeta)$ decreases with the distance to the centre, the equatorial trapping of sub-inertial ($\bar{\nu}>1$) gravito-inertial modes become less and less important.

\section{Asymptotic seismic diagnosis}\label{sect:seismic_diagnosis}
\subsection{Asymptotic period spacing pattern}
\begin{figure}
    \centering
    \resizebox{\hsize}{!}{\includegraphics{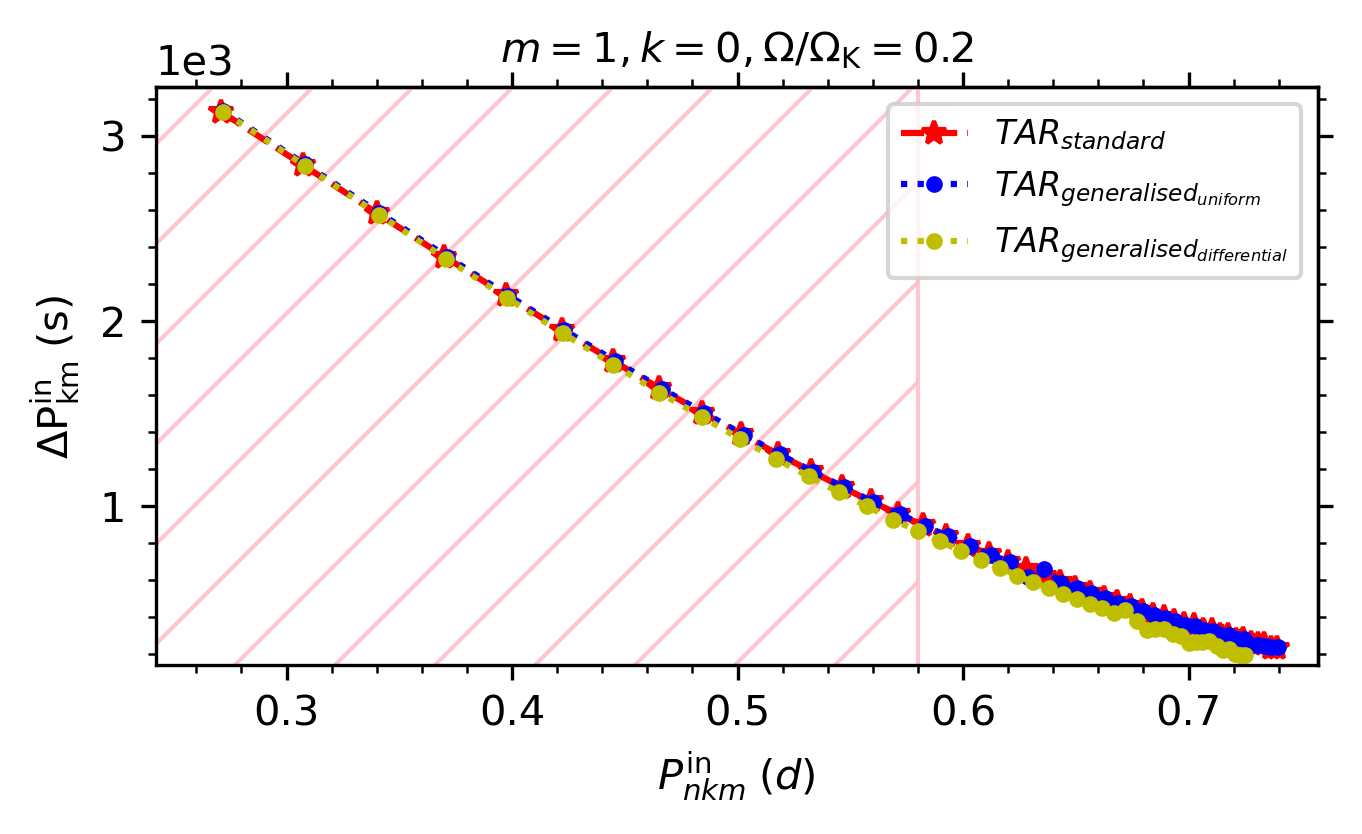}}
    \resizebox{\hsize}{!}{\includegraphics{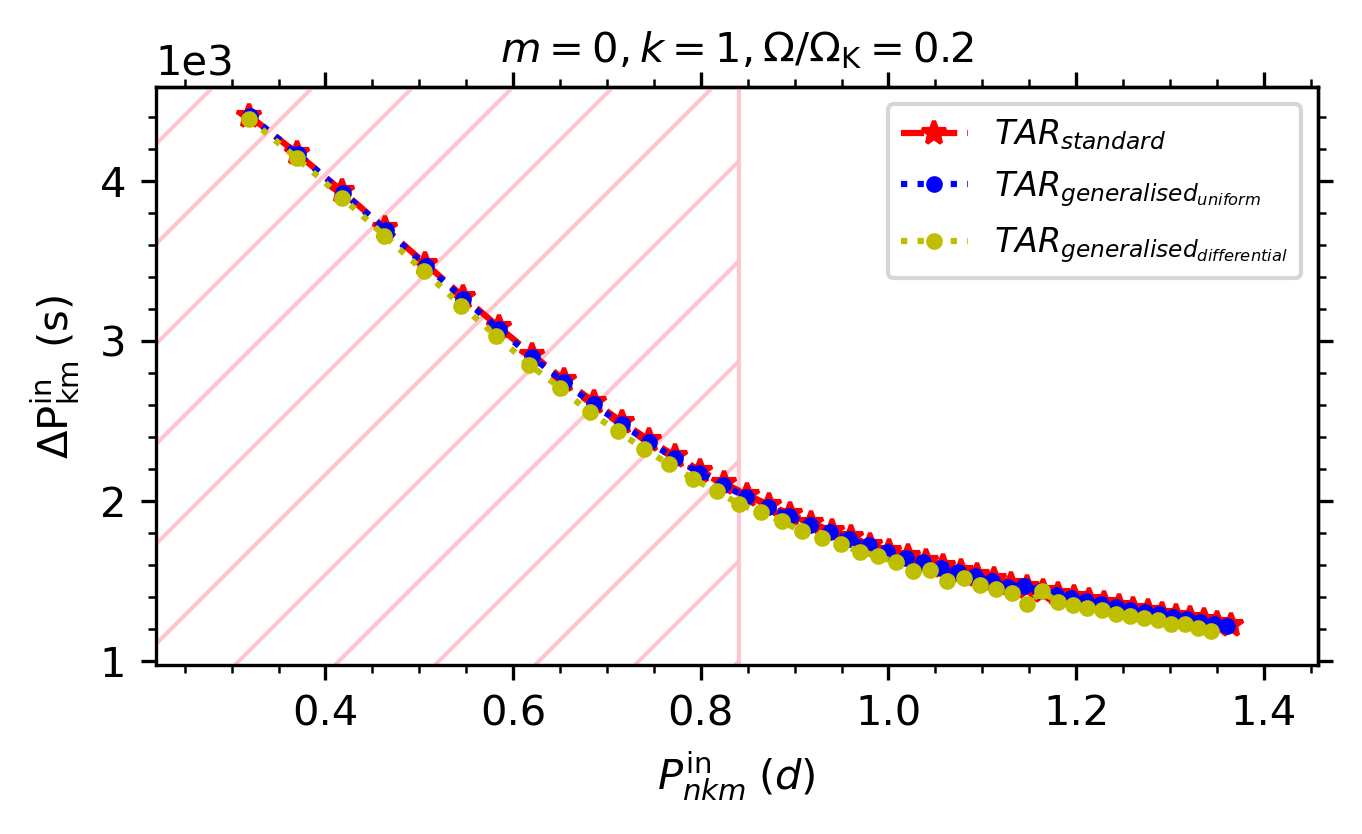}}
    \resizebox{\hsize}{!}{\includegraphics{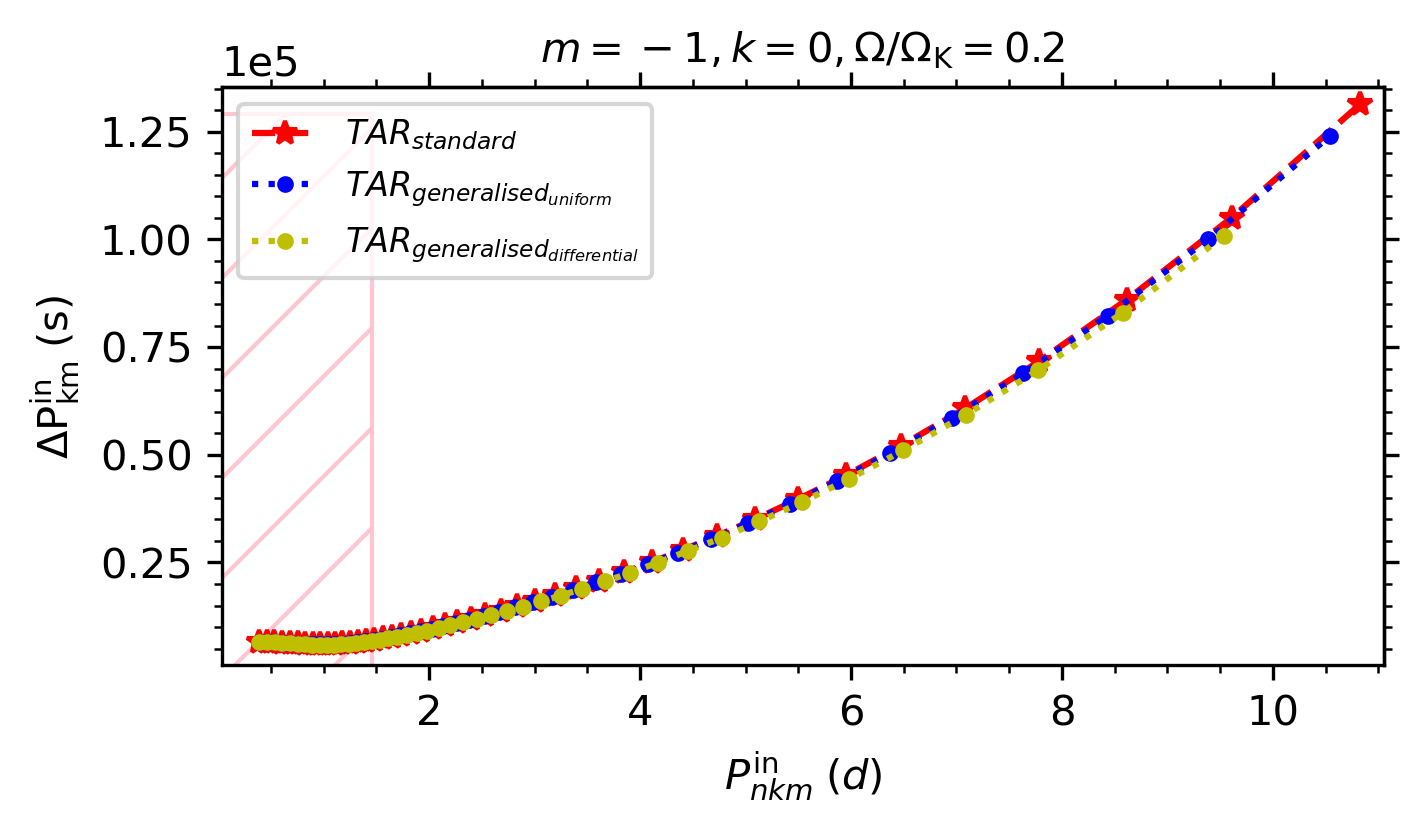}}
    \caption{Period spacing pattern in the inertial frame computed for $\{k=0,m=1\}$ (top) $\{k=1,m=0\}$ (middle),  $\{k=0,m=-1\}$ (bottom) modes at $\Omega/\Omega_{\rm K}=20\%$ in a spherical (red) and a deformed (blue) star  uniformly rotating (at the weighted mean of the rotation rate $\overline{\Omega}=\unit{12.86}{\micro\hertz}$) and in a deformed star differentially rotating (yellow). The hatched pink area indicates where the Cowling approximation is potentially invalid. (The  fluctuations in the period spacing pattern are caused by the numerical noise which is introduced by the numerical derivatives of the mapping with respect to $\zeta$ and $\theta$ used in the resolution of the GLTE).}
    \label{fig:periodspacing}
\end{figure}

In order to compute the period spacing patterns, we adapt the method developed in \cite{Henneco2021} and used in Paper I. First, we calculate for each radial order $n$ of a given mode $(k,m)$, the corresponding asymptotic frequencies in the inertial frame $\omega_{nkm}^{\rm in}$  using Eqs.\;(\ref{eq:doppler_shift}) and (\ref{eq:frequencies}). Then, we can calculate the asymptotic periods in the inertial frame ($P_{nkm}^{\mathrm{in}}=2\pi/ \omega_{nkm}^{\mathrm{in}}$) and the corresponding period spacing which is shown in Fig.\;\ref{fig:periodspacing} for $\{k=0,m=1\}$, $\{k=1,m=0\}$, and $\{k=0,m=-1\}$ modes. The periods represented here are calculated for radial orders between $n=5$ and $n=50$ but the Cowling approximation that we adopted here to develop our formalism is valid only for high radial orders \citep{Bouabid2013,ouazzani2017}. That is why we hatched the region with low radial order ($n<20$), so we can identify where the Cowling approximation is potentially invalid. We find a net decrease in the period spacing $\Delta P_{nkm}^{\mathrm{in}}$
caused by the differential rotation.
The centrifugal and the differential rotation effects are moderately significant when assessed within the validity domain of the generalised TAR nonetheless the global characteristics of the period spacing pattern are conserved. This is in good agreement with the results obtained by \cite{ouazzani2017}. They computed gravito-inertial modes and their period spacing (represented in Fig.\;6 of their article) using the ACOR 2D oscillation code with deformed stellar models (see \cite{Ouazzani2015} for details on their method), which take the centrifugal acceleration into account following the method of deformation of acoustic models by \cite{Roxburgh2006}.  \cite{Ballot2012} also studied the effect of the centrifugal acceleration on the period spacing. They found, for a sectoral mode ($k=0$), a discrepancy between the period spacing (represented in Fig.\;2 of their article) computed using the spherical TAR and the complete computations using TOP with spheroidal models. They demonstrate that this discrepancy originates in the centrifugal distortion of the 2D models. But for other modes, they do not see a considerable effect of the centrifugal deformation on the period spacing pattern in agreement with our results.

Even though the centrifugal deformation effects predicted by the TAR are weak in intermediate-mass stars, they are theoretically detectable for some radial orders of several modes using observations from TESS \citep[Transiting Exoplanet Survey Satellite;][]{ricker2014} and {\it Kepler}  (Paper I).
This is why we do not neglect them and we study the differential rotation and the deformation effects simultaneously.
To quantify the induced variation and evaluate its detectability, we compute the frequency differences.

\subsection{Detectability of the differential rotation effect}
To evaluate the detectability of the differential rotation effect with space-based photometric observations, we compute first the frequency differences $\Delta f_{\rm centrifugal}$ between asymptotic frequencies calculated in the standard TAR ($\rm TAR_s$) and those calculated in the generalised TAR with a uniform rotation ($\rm TAR_{g_{u}}$) and a differential rotation ($\rm TAR_{g_{d}}$) through Eq.\,(\ref{eq:frequencies}):

\begin{equation}
    \Delta f(n) = |f_{\rm TAR_x}(n) - f_{\rm TAR_y}(n) |,
\end{equation}
with $\rm x$ and $\rm y$ two parameters that allow us to choose the effect to be evaluated:
\begin{equation*}
    (\rm x,\rm y) \equiv \left\{
    \begin{array}{ll}
        (\rm g_{u},\rm s) \text{ for the effect of the centrifugal acceleration;} \\
        (\rm g_{u},\rm g_{d}) \text{ for the effect of the differential rotation;} \\
        (\rm g_{d},\rm s) \text{ for the two effects simultaneously.}
    \end{array}
\right.
\end{equation*}

\begin{figure}
    \centering
     \resizebox{\hsize}{!}{\includegraphics{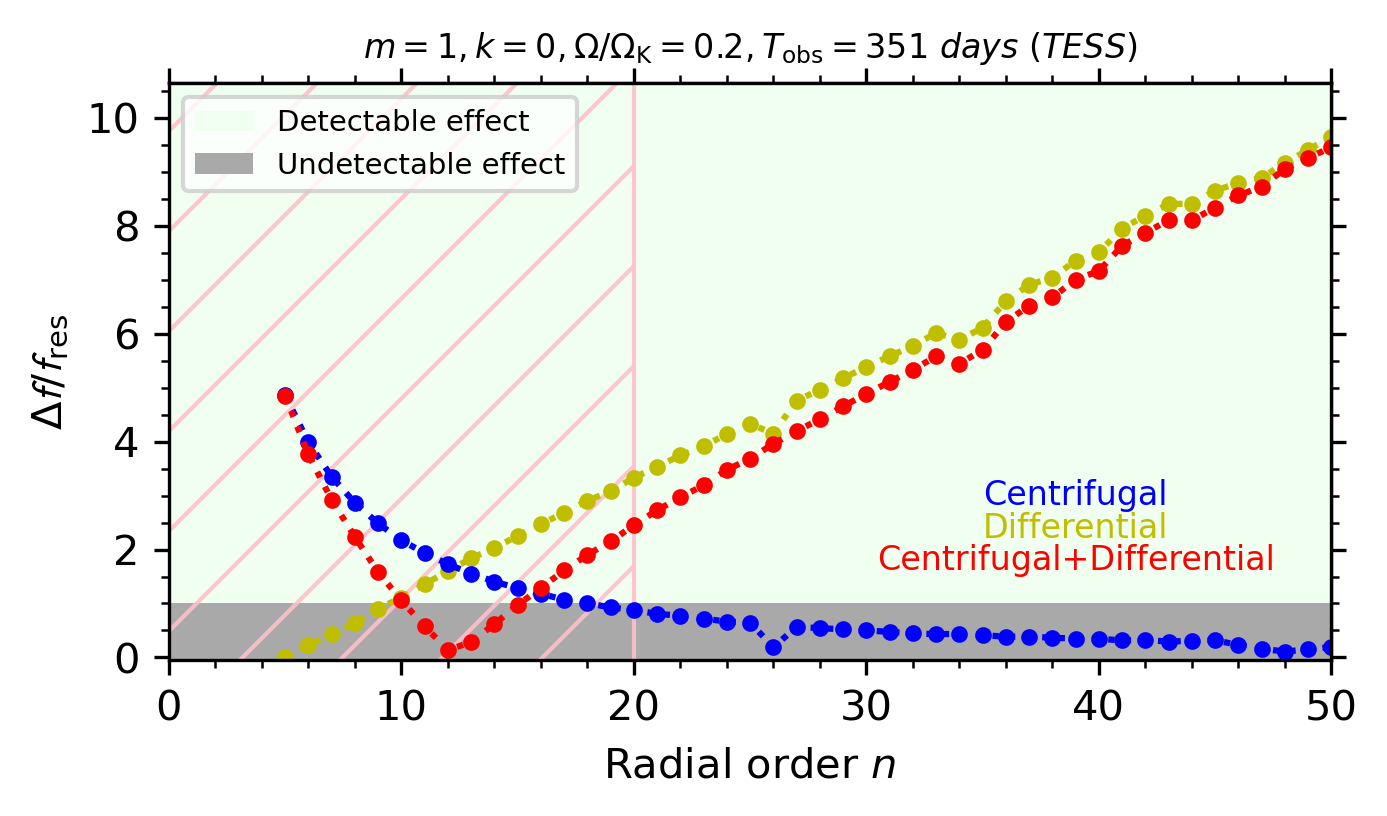}}
     \resizebox{\hsize}{!}{\includegraphics{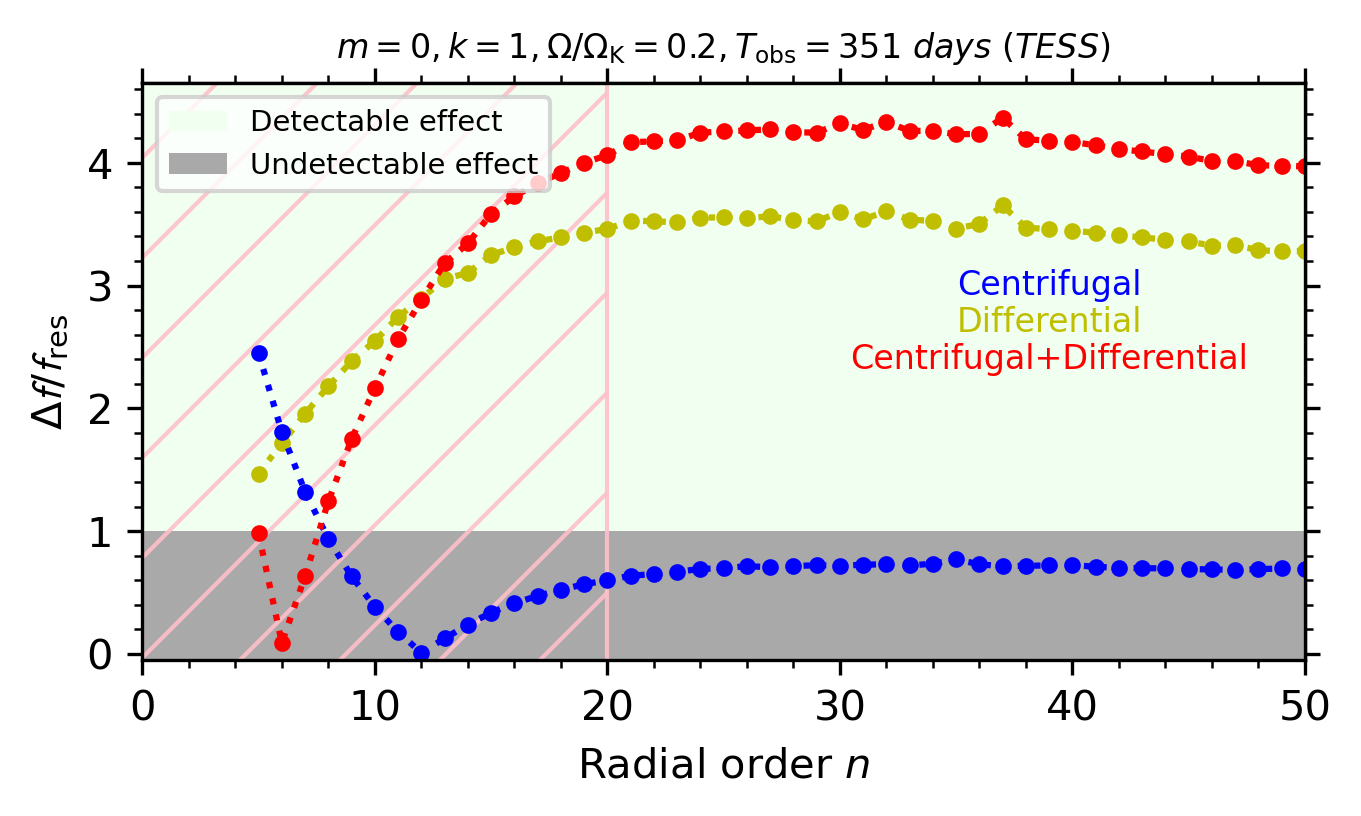}}
     \resizebox{\hsize}{!}{\includegraphics{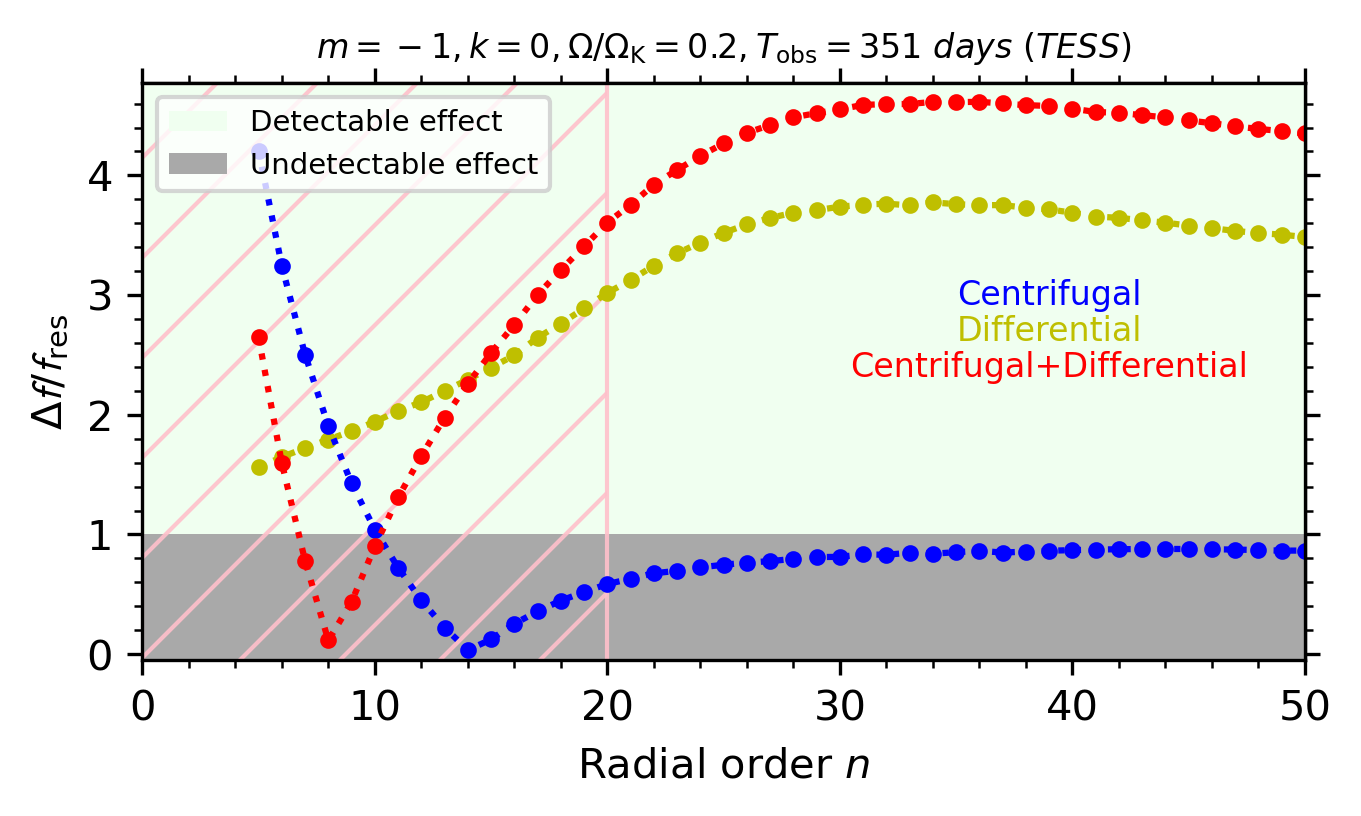}}
    \caption{Detectable radial orders $n$ for $\{k=0,m=1\}$ (top) $\{k=1,m=0\}$ (middle),  $\{k=0,m=-1\}$ (bottom) modes at a rotation rate $\Omega/\Omega_{\mathrm{\rm K}}=20\%$ based on the  frequency resolution of TESS. Blue, yellow and red dots represent respectively the centrifugal effect, the differential rotation effect and the sum of the two effects (the hatched pink area indicates where the Cowling approximation is potentially invalid)}.
    \label{fig:detectability}
\end{figure}

Then, by comparing the obtained frequency differences with the frequency resolutions ($f_{\mathrm{res}}=1/T_{\mathrm{obs}}$) of \textit{Kepler} and TESS light curves covering  quasi-continuously observation times of $T_{\mathrm{obs}} = 4\,$years and $T_{\mathrm{obs}} = 351\,$days, respectively, we can deduce the radial orders $n$'s for which the frequency differences are expected to be theoretically detectable. In Fig.\,\ref{fig:detectability}, we display these results for $\{k=0,m=1\}$, $\{k=1,m=0\}$, and $\{k=0,m=-1\}$ modes rotating at $0.2\,\Omega_{\mathrm{\rm K}}$. In this figure, we show the detectability of the centrifugal effect, the differential rotation effect and of the two effects taken simultaneously into account. Unlike the centrifugal effect, we can see clearly that the effect of the differential rotation for these modes is, theoretically, largely detectable for all radial order higher than $15$ using TESS and \textit{Kepler} observations (not represented here because its observation time is larger than the one of TESS so the detectability using \textit{Kepler} is guaranteed). The detectability of the centrifugal effect is discussed in details in Paper I.

As we pointed out in Paper I, this is a formal evaluation due to the presence of large correlations between the different parameters of our stellar models. This will mask the effect of the centrifugal acceleration and the differential rotation in forward asteroseismic modelling analyses, even when it is theoretically detectable:
\begin{equation}
   \Delta f_{i}>f_{\rm res},\; i\equiv \{\rm centrifugal, \rm differential\}. 
\end{equation}

\section{Evaluation of the terms hierarchy imposed by the TAR within differentially rotating deformed stars}\label{sect:hierarchy_validation}
The hierarchy of terms imposed by the TAR can be summarised by the following frequency hierarchy:
\begin{gather}
    2\Omega\ll N,\label{eq:hier1}\\
    \omega\ll N,\label{eq:hier2}
\end{gather}
which ensures the other hierarchies. Using the Brunt–Väisälä frequency and rotation profiles from ESTER models and using the asymptotic frequencies calculated in Sect.\;\ref{sect:seismic_diagnosis}, we compare these frequencies and we discuss whether the TAR is still valid in differentially rotating deformed stars as in Paper I.

\subsection{The strong stratification assumption: $2\Omega\ll N$}
\begin{figure}
    \centering
    \resizebox{\hsize}{!}{\includegraphics{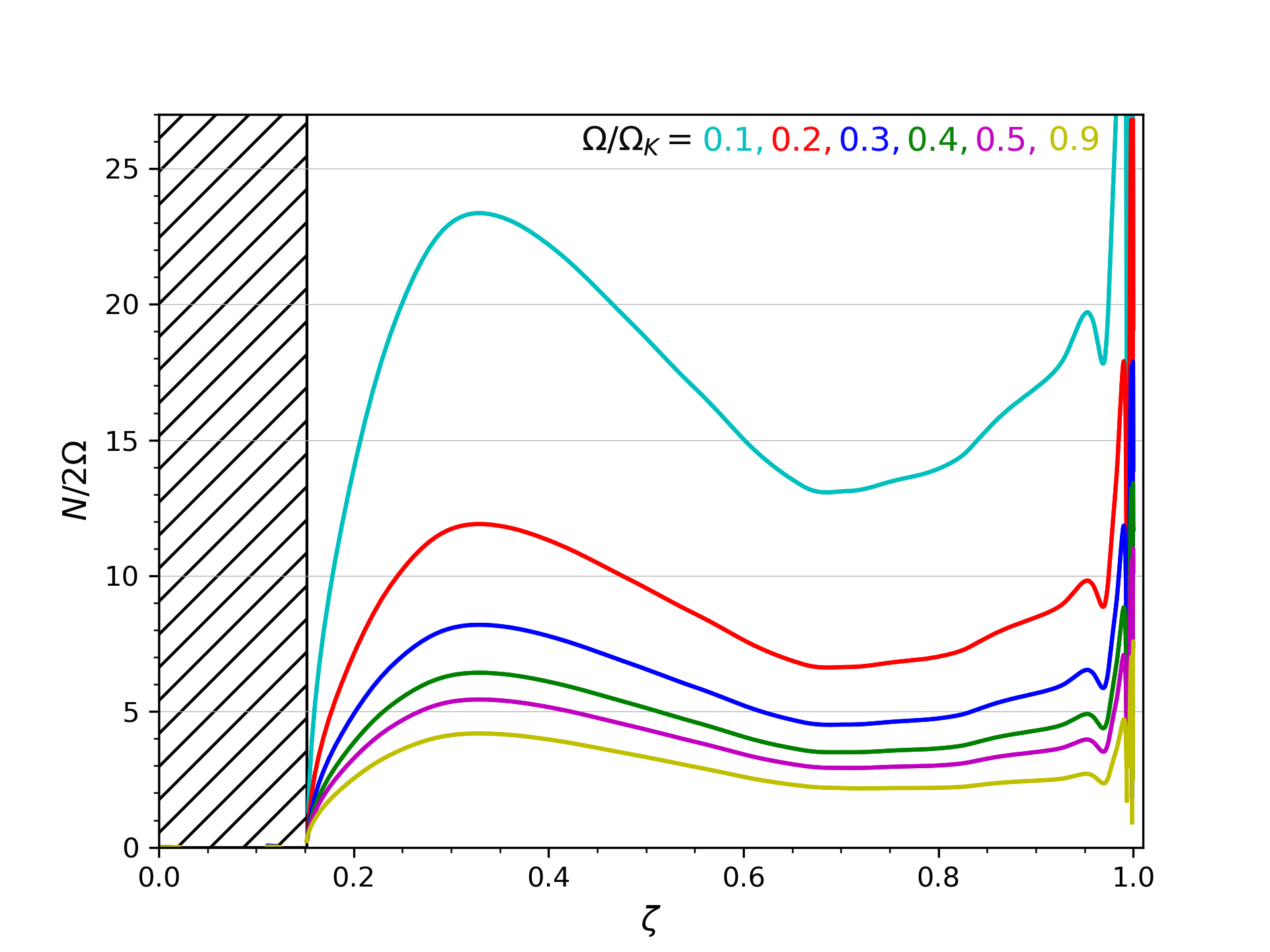}}
    \caption{$N/2\Omega$ term  as a function of the pseudo-radius $\zeta$ at different rotation rates $\Omega/\Omega_{\rm K}$ (the hatched area represents the convective region of the star).}
    \label{fig:tarverifrot}
\end{figure}

We evaluate the term $N/2\Omega$ using $3\,\mathrm{M}_{\odot}$, $X_{\mathrm{c}}=0.7$ ESTER models  for rotation rates $\Omega/\Omega_{\mathrm{\rm K}} \in \left[0.1, 0.9\right]$. As shown in Fig.\,\ref{fig:tarverifrot}, the strong stratification assumption is valid only in the radiative zone away from the border between the convective core and the radiative envelope ($\zeta \ge 0.2 $) at $\Omega/\Omega_{\mathrm{\rm K}}\le 0.2$. Beyond this critical value, the Brunt–Väisälä frequency and the frequency of rotation have a close order of magnitude so the strong stratification approximation is no longer valid.

\subsection{The low frequency assumption: $\omega\ll N$}
\begin{figure}
    \centering
    \resizebox{\hsize}{!}{\includegraphics{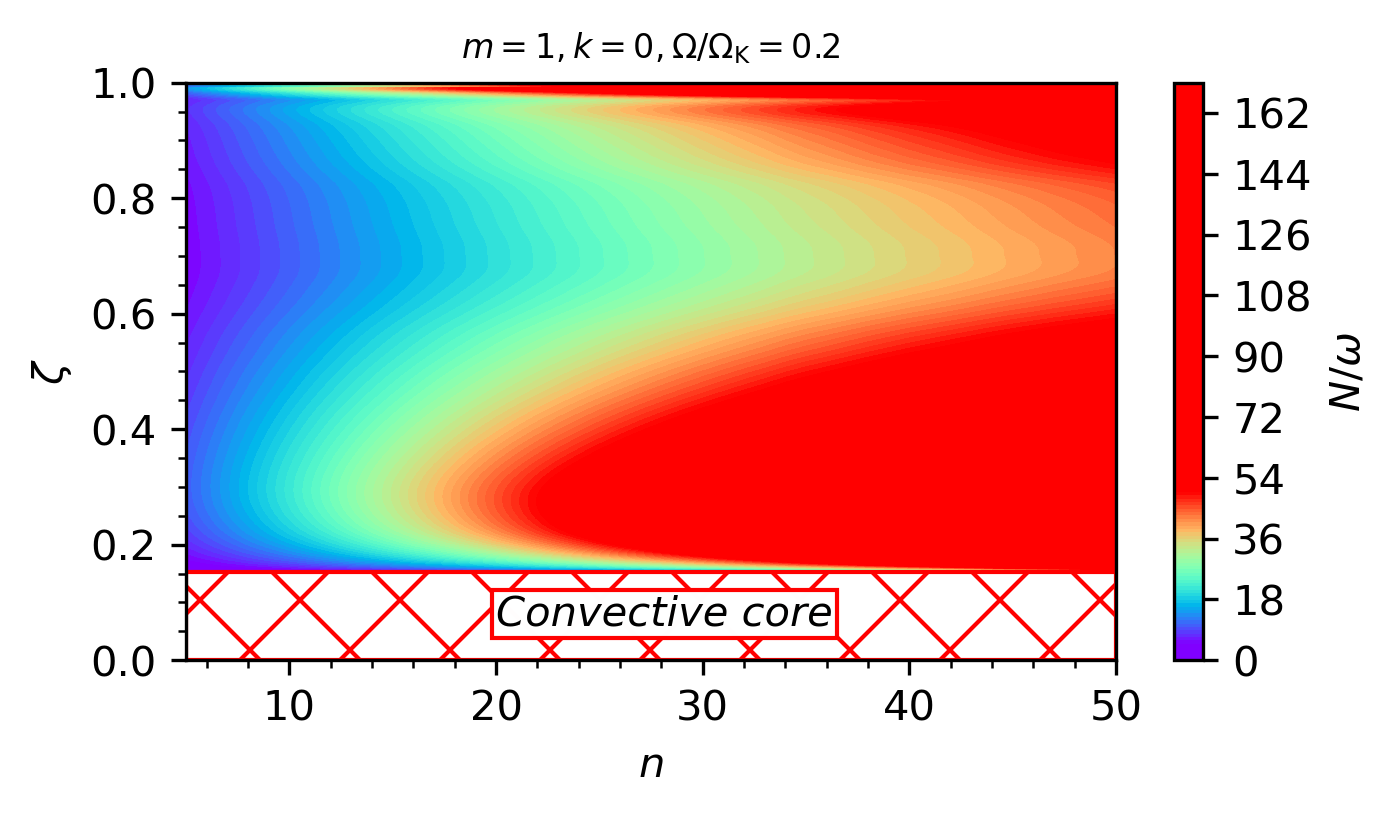}}
    \resizebox{\hsize}{!}{\includegraphics{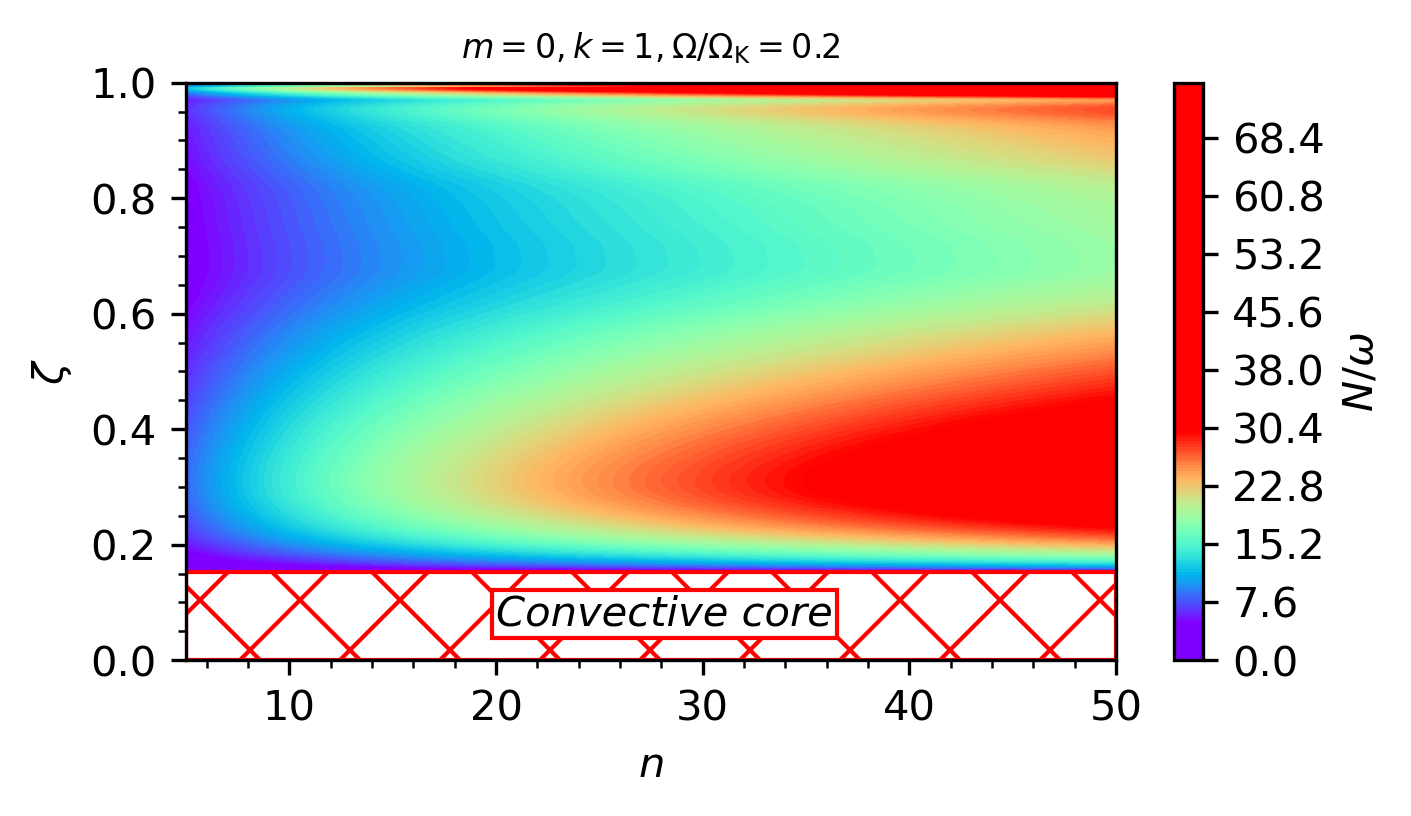}}
    \resizebox{\hsize}{!}{\includegraphics{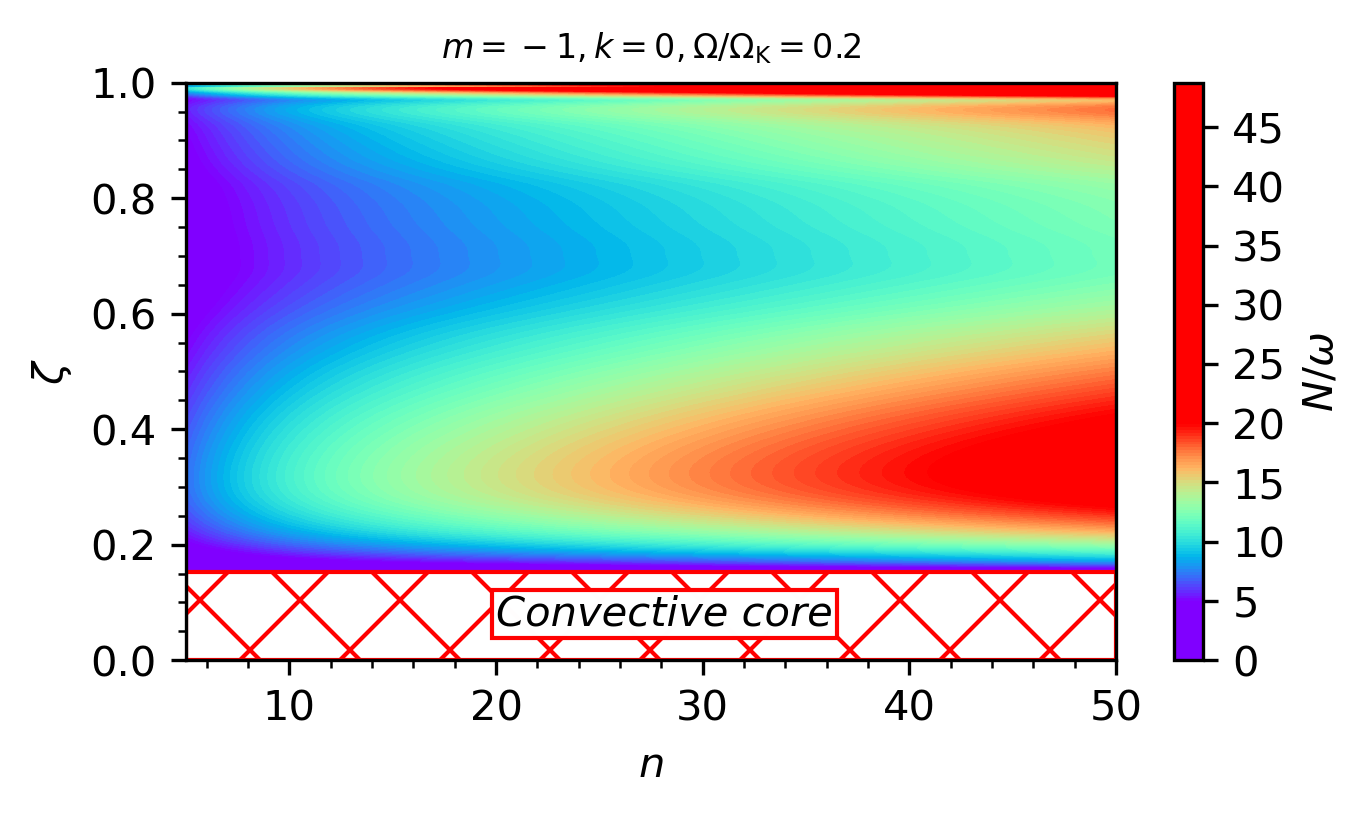}}
    \caption{$N/\omega$ term as a function of the pseudo-radius $\zeta$ and the radial order $n$ for the $\{k=0,m=1\}$ (top), $\{k=1,m=0\}$ (middle) and $\{k=0,m=-1\}$ (bottom) modes at $\Omega/\Omega_{\rm K}=20\%$.}
    \label{fig:tarveriffreq}
\end{figure}

We evaluate the term $N/\omega$  using  the asymptotic frequencies for $\{k=0,m=1\}$, $\{k=1,m=0\}$,  and $\{k=0,m=-1\}$
modes calculated in Sect.\;\ref{sect:seismic_diagnosis}. As shown in Fig.\,\ref{fig:tarveriffreq}, the low frequency assumption is valid in all the space domain for the $\{k=0,m=1\}$ mode only for high radial orders ($n>20$). Below this critical value, the Brunt–Väisälä frequency and the wave frequency can have a very close order of magnitude near the transition layer between the convective core and the radiative envelope. But far from this interface, the low frequency approximation is valid for all the modes.

\section{Discussion and conclusions} \label{sect:conclusion}
This work is a continuation of our previous study where we generalised the TAR in the case of strongly deformed, rapidly and uniformly rotating stars and planets. In this perspective, we study the possibility of carrying out a new generalisation of the TAR that abandons the assumption of  uniform rotation and takes into account the radial and latitudinal differential rotation.
We approached this exploration by deriving the generalised Laplace Tidal equation in a spheroidal coordinate system. We relied mainly on two assumptions that will define the validity domain of the generalised TAR (Paper I). The equation that we derive has a similar form to the one that is obtained when the TAR is applied to weakly rotating spherical stars \citep{lee+saio1997}, differentially rotating spherical stars \citep{mathis2009,vanreeth2018}, moderately rotating weakly deformed stars \citep{mathis+prat2019, Henneco2021} and uniformly rotating strongly deformed stars (Paper I). So with this new formalism, we can study GIWs in the radiative region of all types of stars and planets.

We apply this general formalism to rapidly and differentially rotating early-type stars using 2D ESTER models $(\mathrm{M}=3 \mathrm{M}_{\odot},~X_{\rm c} = 0.7)$ where we found that the signature of the differential rotation effect in the period spacing patterns is stronger than the signature of the centrifugal effect for the $\{k=0,m=1\}$ mode at $\Omega/\Omega_{\rm K}=20\%$, typically up to a factor twenty. This is caused essentially by the presence of a strong pseudo-radial $\Omega$-gradients in stellar interior of early-type stars. 
This new generalisation of the TAR can be used to study the dissipation of stellar and planetary tides induced by low-frequency GIWs in rapidly and differentially rotating deformed stars and planets and the angular momentum transport with a formalism that can be directly implemented in ESTER models.

The next step will be the inclusion of the magnetic field in a non-perturbative way within the generalised TAR formalism. So far, \cite{prat2019,prat2020} and \cite{VanBeeck2020} have focused on the case where magnetic fields are weak enough to be treated within a perturbative treatment to study the effects of a magnetic field on the seismic parameters of $\rm g$ modes which become magneto-gravito inertial modes \citep{Mathis+deBrye2011, Mathis+deBrye2012}. So a possible follow-up of this work (Paper III) is to take into account the stellar magnetic fields and generalise the TAR framework to the case of differentially rotating strongly deformed magnetic stars.

\begin{acknowledgements}
We thank the referee for their constructive comments and suggestions that allow us to improve our article. We thank the ESTER code developers for their efforts and for making their codes publicly available. We are also grateful to Pr.~C.Aerts who kindly commented on the manuscript and suggested improvements. H.D., V.P. and S.M. acknowledge support from the European Research Council through ERC grant SPIRE 647383. H.D. and S.M. acknowledge support from the CNES PLATO grant at CEA/DAp. TVR gratefully acknowledges support from the Research Foundation Flanders (FWO) under grant agreement No. 12ZB620N.
\end{acknowledgements}

\bibliographystyle{aa}
\bibliography{bibliography}

\begin{thebibliography}{108}
\expandafter\ifx\csname natexlab\endcsname\relax\def\natexlab#1{#1}\fi

\bibitem[{Aerts(2021)}]{aerts2021}
Aerts, C. 2021, Rev. Mod. Phys., 93, 015001

\bibitem[{{Aerts} {et~al.}(2019){Aerts}, {Mathis}, \& {Rogers}}]{aerts2019}
{Aerts}, C., {Mathis}, S., \& {Rogers}, T.~M. 2019, \araa, 57, 35

\bibitem[{{Ballot} {et~al.}(2012){Ballot}, {Ligni{\`e}res}, {Prat}, {Reese}, \&
  {Rieutord}}]{Ballot2012}
{Ballot}, J., {Ligni{\`e}res}, F., {Prat}, V., {Reese}, D.~R., \& {Rieutord},
  M. 2012, in Astronomical Society of the Pacific Conference Series, Vol. 462,
  Progress in Solar/Stellar Physics with Helio- and Asteroseismology, ed.
  H.~{Shibahashi}, M.~{Takata}, \& A.~E. {Lynas-Gray}, 389

\bibitem[{{Ballot} {et~al.}(2010){Ballot}, {Ligni{\`e}res}, {Reese}, \&
  {Rieutord}}]{Ballot2010}
{Ballot}, J., {Ligni{\`e}res}, F., {Reese}, D.~R., \& {Rieutord}, M. 2010,
  \aap, 518, A30

\bibitem[{{Beck} {et~al.}(2014){Beck}, {Hambleton}, {Vos}, {Kallinger},
  {Bloemen}, {Tkachenko}, {Garc{\'\i}a}, {{\O}stensen}, {Aerts}, {Kurtz}, {De
  Ridder}, {Hekker}, {Pavlovski}, {Mathur}, {De Smedt}, {Derekas}, {Corsaro},
  {Mosser}, {Van Winckel}, {Huber}, {Degroote}, {Davies}, {Pr{\v{s}}a},
  {Debosscher}, {Elsworth}, {Nemeth}, {Siess}, {Schmid}, {P{\'a}pics}, {de
  Vries}, {van Marle}, {Marcos-Arenal}, \& {Lobel}}]{Beck2014RG}
{Beck}, P.~G., {Hambleton}, K., {Vos}, J., {et~al.} 2014, \aap, 564, A36

\bibitem[{{Beck} {et~al.}(2018){Beck}, {Kallinger}, {Pavlovski}, {Palacios},
  {Tkachenko}, {Mathis}, {Garc{\'\i}a}, {Corsaro}, {Johnston}, {Mosser},
  {Ceillier}, {do Nascimento}, \& {Raskin}}]{Beck2018RG}
{Beck}, P.~G., {Kallinger}, T., {Pavlovski}, K., {et~al.} 2018, \aap, 612, A22

\bibitem[{{Beck} {et~al.}(2012){Beck}, {Montalban}, {Kallinger}, {De Ridder},
  {Aerts}, {Garc{\'\i}a}, {Hekker}, {Dupret}, {Mosser}, {Eggenberger},
  {Stello}, {Elsworth}, {Frandsen}, {Carrier}, {Hillen}, {Gruberbauer},
  {Christensen-Dalsgaard}, {Miglio}, {Valentini}, {Bedding}, {Kjeldsen},
  {Girouard}, {Hall}, \& {Ibrahim}}]{Beck2012RG}
{Beck}, P.~G., {Montalban}, J., {Kallinger}, T., {et~al.} 2012, \nat, 481, 55

\bibitem[{{Bedding} {et~al.}(2015){Bedding}, {Murphy}, {Colman}, \&
  {Kurtz}}]{Beddingetal2015Gamma}
{Bedding}, T.~R., {Murphy}, S.~J., {Colman}, I.~L., \& {Kurtz}, D.~W. 2015, in
  European Physical Journal Web of Conferences, Vol. 101, European Physical
  Journal Web of Conferences, 01005

\bibitem[{{Bildsten} {et~al.}(1996){Bildsten}, {Ushomirsky}, \&
  {Cutler}}]{Bildsten1996}
{Bildsten}, L., {Ushomirsky}, G., \& {Cutler}, C. 1996, \apj, 460, 827

\bibitem[{Bonazzola {et~al.}(1998)Bonazzola, Gourgoulhon, \&
  Marck}]{bonazzola1998}
Bonazzola, S., Gourgoulhon, E., \& Marck, J.-A. 1998, Phys. Rev. D, 58, 104020

\bibitem[{{Borucki} {et~al.}(2009){Borucki}, {Koch}, {Batalha}, {Caldwell},
  {Christensen-Dalsgaard}, {Cochran}, {Dunham}, {Gautier}, {Geary},
  {Gilliland}, {Jenkins}, {Kjeldsen}, {Lissauer}, \& {Rowe}}]{borucki2009}
{Borucki}, W., {Koch}, D., {Batalha}, N., {et~al.} 2009, in IAU Symposium, Vol.
  253, Transiting Planets, ed. F.~{Pont}, D.~{Sasselov}, \& M.~J. {Holman},
  289--299

\bibitem[{{Bouabid} {et~al.}(2013){Bouabid}, {Dupret}, {Salmon},
  {Montalb{\'a}n}, {Miglio}, \& {Noels}}]{Bouabid2013}
{Bouabid}, M.~P., {Dupret}, M.~A., {Salmon}, S., {et~al.} 2013, \mnras, 429,
  2500

\bibitem[{{Buysschaert} {et~al.}(2018){Buysschaert}, {Aerts}, {Bowman},
  {Johnston}, {Van Reeth}, {Pedersen}, {Mathis}, \&
  {Neiner}}]{Buysschaert2018B}
{Buysschaert}, B., {Aerts}, C., {Bowman}, D.~M., {et~al.} 2018, \aap, 616, A148

\bibitem[{{Cantiello} {et~al.}(2014){Cantiello}, {Mankovich}, {Bildsten},
  {Christensen-Dalsgaard}, \& {Paxton}}]{Cantiello2014}
{Cantiello}, M., {Mankovich}, C., {Bildsten}, L., {Christensen-Dalsgaard}, J.,
  \& {Paxton}, B. 2014, \apj, 788, 93

\bibitem[{{Ceillier} {et~al.}(2013){Ceillier}, {Eggenberger}, {Garc{\'\i}a}, \&
  {Mathis}}]{Ceillier2013}
{Ceillier}, T., {Eggenberger}, P., {Garc{\'\i}a}, R.~A., \& {Mathis}, S. 2013,
  \aap, 555, A54

\bibitem[{{Charbonnel} {et~al.}(2013){Charbonnel}, {Decressin}, {Amard},
  {Palacios}, \& {Talon}}]{charbonnel2013}
{Charbonnel}, C., {Decressin}, T., {Amard}, L., {Palacios}, A., \& {Talon}, S.
  2013, \aap, 554, A40

\bibitem[{Christensen-Dalsgaard(1997)}]{Christensen1997}
Christensen-Dalsgaard, J. 1997

\bibitem[{{Christophe} {et~al.}(2018){Christophe}, {Ouazzani}, {Ballot},
  {Marques}, {Goupil}, {Antoci}, \& {Salmon}}]{Christophe2018}
{Christophe}, S., {Ouazzani}, R.~M., {Ballot}, J., {et~al.} 2018, in SF2A-2018:
  Proceedings of the Annual meeting of the French Society of Astronomy and
  Astrophysics, ed. P.~{Di Matteo}, F.~{Billebaud}, F.~{Herpin}, N.~{Lagarde},
  J.~B. {Marquette}, A.~{Robin}, \& O.~{Venot}, Di

\bibitem[{{Cowling}(1941)}]{cowling1941}
{Cowling}, T.~G. 1941, \mnras, 101, 367

\bibitem[{{Decressin} {et~al.}(2009){Decressin}, {Mathis}, {Palacios}, {Siess},
  {Talon}, {Charbonnel}, \& {Zahn}}]{Decressin2009}
{Decressin}, T., {Mathis}, S., {Palacios}, A., {et~al.} 2009, \aap, 495, 271

\bibitem[{{Deheuvels} {et~al.}(2015){Deheuvels}, {Ballot}, {Beck}, {Mosser},
  {{\O}stensen}, {Garc{\'\i}a}, \& {Goupil}}]{Deheuvels2015RG}
{Deheuvels}, S., {Ballot}, J., {Beck}, P.~G., {et~al.} 2015, \aap, 580, A96

\bibitem[{{Deheuvels} {et~al.}(2020){Deheuvels}, {Ballot}, {Eggenberger},
  {Spada}, {Noll}, \& {den Hartogh}}]{Deheuvels2020RG}
{Deheuvels}, S., {Ballot}, J., {Eggenberger}, P., {et~al.} 2020, \aap, 641,
  A117

\bibitem[{{Deheuvels} {et~al.}(2014){Deheuvels}, {Do{\u{g}}an}, {Goupil},
  {Appourchaux}, {Benomar}, {Bruntt}, {Campante}, {Casagrande}, {Ceillier},
  {Davies}, {De Cat}, {Fu}, {Garc{\'\i}a}, {Lobel}, {Mosser}, {Reese},
  {Regulo}, {Schou}, {Stahn}, {Thygesen}, {Yang}, {Chaplin},
  {Christensen-Dalsgaard}, {Eggenberger}, {Gizon}, {Mathis},
  {Molenda-{\.Z}akowicz}, \& {Pinsonneault}}]{Deheuvels2014RG}
{Deheuvels}, S., {Do{\u{g}}an}, G., {Goupil}, M.~J., {et~al.} 2014, \aap, 564,
  A27

\bibitem[{{Deheuvels} {et~al.}(2012){Deheuvels}, {Garc{\'\i}a}, {Chaplin},
  {Basu}, {Antia}, {Appourchaux}, {Benomar}, {Davies}, {Elsworth}, {Gizon},
  {Goupil}, {Reese}, {Regulo}, {Schou}, {Stahn}, {Casagrande},
  {Christensen-Dalsgaard}, {Fischer}, {Hekker}, {Kjeldsen}, {Mathur}, {Mosser},
  {Pinsonneault}, {Valenti}, {Christiansen}, {Kinemuchi}, \&
  {Mullally}}]{Deheuvels2012RG}
{Deheuvels}, S., {Garc{\'\i}a}, R.~A., {Chaplin}, W.~J., {et~al.} 2012, \apj,
  756, 19

\bibitem[{{Dhouib} {et~al.}(2021){Dhouib}, {Prat}, {Van Reeth}, \&
  {Mathis}}]{dhouib2021}
{Dhouib}, H., {Prat}, V., {Van Reeth}, T., \& {Mathis}, S. 2021, A\&A, 652,
  A154

\bibitem[{{Di Mauro} {et~al.}(2016){Di Mauro}, {Ventura}, {Cardini}, {Stello},
  {Christensen-Dalsgaard}, {Dziembowski}, {Patern{\`o}}, {Beck}, {Bloemen},
  {Davies}, {De Smedt}, {Elsworth}, {Garc{\'\i}a}, {Hekker}, {Mosser}, \&
  {Tkachenko}}]{DiMauro2016RG}
{Di Mauro}, M.~P., {Ventura}, R., {Cardini}, D., {et~al.} 2016, \apj, 817, 65

\bibitem[{{Dintrans} \& {Rieutord}(2000)}]{Dintrans2000}
{Dintrans}, B. \& {Rieutord}, M. 2000, \aap, 354, 86

\bibitem[{{Dintrans} {et~al.}(1999){Dintrans}, {Rieutord}, \&
  {Valdettaro}}]{Dintrans1999}
{Dintrans}, B., {Rieutord}, M., \& {Valdettaro}, L. 1999, Journal of Fluid
  Mechanics, 398, 271

\bibitem[{{Eckart}(1960)}]{eckart1960}
{Eckart}, C. 1960, {Hydrodynamics of Oceans and Atmospheres} (Pergamon Press
  (Oxford))

\bibitem[{{Eggenberger} {et~al.}(2012){Eggenberger}, {Montalb{\'a}n}, \&
  {Miglio}}]{Eggenberger2012}
{Eggenberger}, P., {Montalb{\'a}n}, J., \& {Miglio}, A. 2012, \aap, 544, L4

\bibitem[{{Espinosa Lara} \& {Rieutord}(2013)}]{espinosa+rieutord2013}
{Espinosa Lara}, F. \& {Rieutord}, M. 2013, A\&A, 552, A35

\bibitem[{Fr{\"o}man \& Fr{\"o}man(1965)}]{froman2005}
Fr{\"o}man, N. \& Fr{\"o}man, P.~O. 1965, JWKB approximation : contributions to
  the theory (Amsterdam: North-Holland Publishing Company)

\bibitem[{{Gallet} \& {Bouvier}(2015)}]{Gallet+Bouvier2015}
{Gallet}, F. \& {Bouvier}, J. 2015, \aap, 577, A98

\bibitem[{{Gehan} {et~al.}(2018){Gehan}, {Mosser}, {Michel}, {Samadi}, \&
  {Kallinger}}]{Gehan2018RG}
{Gehan}, C., {Mosser}, B., {Michel}, E., {Samadi}, R., \& {Kallinger}, T. 2018,
  \aap, 616, A24

\bibitem[{Gough(1993)}]{gough1993}
Gough, D. 1993, Les Houches Session XLVII, ed. J.-P. Zahn, \& J. Zinn-Justin
  (Amsterdam: Elsevier), 399

\bibitem[{{Guo} {et~al.}(2017){Guo}, {Gies}, \& {Matson}}]{Guoetal2017Fbin}
{Guo}, Z., {Gies}, D.~R., \& {Matson}, R.~A. 2017, \apj, 851, 39

\bibitem[{{Henneco} {et~al.}(2021){Henneco}, {Van Reeth}, {Prat}, {Mathis},
  {Mombarg}, \& {Aerts}}]{Henneco2021}
{Henneco}, J., {Van Reeth}, T., {Prat}, V., {et~al.} 2021, \aap, 648, A97

\bibitem[{{Kallinger} {et~al.}(2017){Kallinger}, {Weiss}, {Beck}, {Pigulski},
  {Kuschnig}, {Tkachenko}, {Pakhomov}, {Ryabchikova}, {L{\"u}ftinger}, {Palle},
  {Semenko}, {Handler}, {Koudelka}, {Matthews}, {Moffat}, {Pablo}, {Popowicz},
  {Rucinski}, {Wade}, \& {Zwintz}}]{Kallingeretal2017Bbin}
{Kallinger}, T., {Weiss}, W.~W., {Beck}, P.~G., {et~al.} 2017, \aap, 603, A13

\bibitem[{{Keen} {et~al.}(2015){Keen}, {Bedding}, {Murphy}, {Schmid}, {Aerts},
  {Tkachenko}, {Ouazzani}, \& {Kurtz}}]{Keen2015Gammabin}
{Keen}, M.~A., {Bedding}, T.~R., {Murphy}, S.~J., {et~al.} 2015, \mnras, 454,
  1792

\bibitem[{{Kurtz} {et~al.}(2014){Kurtz}, {Saio}, {Takata}, {Shibahashi},
  {Murphy}, \& {Sekii}}]{Kurtz2014A}
{Kurtz}, D.~W., {Saio}, H., {Takata}, M., {et~al.} 2014, \mnras, 444, 102

\bibitem[{{Lee} \& {Baraffe}(1995)}]{lee+baraffe1995}
{Lee}, U. \& {Baraffe}, I. 1995, \aap, 301, 419

\bibitem[{{Lee} \& {Saio}(1987)}]{lee+saio1987}
{Lee}, U. \& {Saio}, H. 1987, \mnras, 225, 643

\bibitem[{{Lee} \& {Saio}(1997)}]{lee+saio1997}
{Lee}, U. \& {Saio}, H. 1997, \apj, 491, 839

\bibitem[{{Li} {et~al.}(2019){Li}, {Van Reeth}, {Bedding}, {Murphy}, \&
  {Antoci}}]{Li2019Gamma}
{Li}, G., {Van Reeth}, T., {Bedding}, T.~R., {Murphy}, S.~J., \& {Antoci}, V.
  2019, \mnras, 487, 782

\bibitem[{{Li} {et~al.}(2020){Li}, {Van Reeth}, {Bedding}, {Murphy}, {Antoci},
  {Ouazzani}, \& {Barbara}}]{Li2020Gamma}
{Li}, G., {Van Reeth}, T., {Bedding}, T.~R., {et~al.} 2020, \mnras, 491, 3586

\bibitem[{{Ligni{\`e}res} {et~al.}(2006){Ligni{\`e}res}, {Rieutord}, \&
  {Reese}}]{lignieres2006}
{Ligni{\`e}res}, F., {Rieutord}, M., \& {Reese}, D. 2006, \aap, 455, 607

\bibitem[{{Maeder}(2003)}]{Maeder2003}
{Maeder}, A. 2003, \aap, 399, 263

\bibitem[{{Maeder}(2009)}]{Maeder2009}
{Maeder}, A. 2009, {Physics, Formation and Evolution of Rotating Stars}

\bibitem[{{Maeder} \& {Meynet}(2000)}]{Maeder&Meynet2000}
{Maeder}, A. \& {Meynet}, G. 2000, \aap, 361, 159

\bibitem[{{Maeder} \& {Zahn}(1998)}]{Maeder+zahn1998}
{Maeder}, A. \& {Zahn}, J.-P. 1998, \aap, 334, 1000

\bibitem[{{Marques} {et~al.}(2013){Marques}, {Goupil}, {Lebreton}, {Talon},
  {Palacios}, {Belkacem}, {Ouazzani}, {Mosser}, {Moya}, {Morel}, {Pichon},
  {Mathis}, {Zahn}, {Turck-Chi{\`e}ze}, \& {Nghiem}}]{Marques2013}
{Marques}, J.~P., {Goupil}, M.~J., {Lebreton}, Y., {et~al.} 2013, \aap, 549,
  A74

\bibitem[{{Mathis}(2009)}]{mathis2009}
{Mathis}, S. 2009, A\&A, 506, 811

\bibitem[{{Mathis}(2013)}]{Mathis2013}
{Mathis}, S. 2013, {Transport Processes in Stellar Interiors}, ed. M.~{Goupil},
  K.~{Belkacem}, C.~{Neiner}, F.~{Ligni{\`e}res}, \& J.~J. {Green}, Vol. 865,
  23

\bibitem[{{Mathis} \& {de Brye}(2011)}]{Mathis+deBrye2011}
{Mathis}, S. \& {de Brye}, N. 2011, \aap, 526, A65

\bibitem[{{Mathis} \& {de Brye}(2012)}]{Mathis+deBrye2012}
{Mathis}, S. \& {de Brye}, N. 2012, \aap, 540, A37

\bibitem[{{Mathis} {et~al.}(2014){Mathis}, {Neiner}, \& {Tran Minh}}]{MNT2014}
{Mathis}, S., {Neiner}, C., \& {Tran Minh}, N. 2014, \aap, 565, A47

\bibitem[{{Mathis} {et~al.}(2004){Mathis}, {Palacios}, \& {Zahn}}]{Mathis2004}
{Mathis}, S., {Palacios}, A., \& {Zahn}, J.~P. 2004, \aap, 425, 243

\bibitem[{{Mathis} \& {Prat}(2019)}]{mathis+prat2019}
{Mathis}, S. \& {Prat}, V. 2019, A\&A, 631, A26

\bibitem[{{Mathis} {et~al.}(2018){Mathis}, {Prat}, {Amard}, {Charbonnel},
  {Palacios}, {Lagarde}, \& {Eggenberger}}]{mathis2018}
{Mathis}, S., {Prat}, V., {Amard}, L., {et~al.} 2018, \aap, 620, A22

\bibitem[{{Mathis} \& {Zahn}(2004)}]{mathis+zahn2004}
{Mathis}, S. \& {Zahn}, J.~P. 2004, \aap, 425, 229

\bibitem[{{Mestel} \& {Weiss}(1987)}]{Mestel+Weiss1987}
{Mestel}, L. \& {Weiss}, N.~O. 1987, \mnras, 226, 123

\bibitem[{{Mirouh} {et~al.}(2016){Mirouh}, {Baruteau}, {Rieutord}, \&
  {Ballot}}]{Mirouh2016}
{Mirouh}, G.~M., {Baruteau}, C., {Rieutord}, M., \& {Ballot}, J. 2016, Journal
  of Fluid Mechanics, 800, 213

\bibitem[{{Mombarg} {et~al.}(2019){Mombarg}, {Van Reeth}, {Pedersen},
  {Molenberghs}, {Bowman}, {Johnston}, {Tkachenko}, \& {Aerts}}]{Mombarg2019F}
{Mombarg}, J.~S.~G., {Van Reeth}, T., {Pedersen}, M.~G., {et~al.} 2019, \mnras,
  485, 3248

\bibitem[{{Moravveji} {et~al.}(2016){Moravveji}, {Townsend}, {Aerts}, \&
  {Mathis}}]{Moravveji2016}
{Moravveji}, E., {Townsend}, R. H.~D., {Aerts}, C., \& {Mathis}, S. 2016, \apj,
  823, 130

\bibitem[{{Mosser} {et~al.}(2012){Mosser}, {Goupil}, {Belkacem}, {Marques},
  {Beck}, {Bloemen}, {De Ridder}, {Barban}, {Deheuvels}, {Elsworth}, {Hekker},
  {Kallinger}, {Ouazzani}, {Pinsonneault}, {Samadi}, {Stello}, {Garc{\'\i}a},
  {Klaus}, {Li}, {Mathur}, \& {Morris}}]{Mosser2012RG}
{Mosser}, B., {Goupil}, M.~J., {Belkacem}, K., {et~al.} 2012, \aap, 548, A10

\bibitem[{{Murphy} {et~al.}(2016){Murphy}, {Fossati}, {Bedding}, {Saio},
  {Kurtz}, {Grassitelli}, \& {Wang}}]{Murphy2016Gamma}
{Murphy}, S.~J., {Fossati}, L., {Bedding}, T.~R., {et~al.} 2016, \mnras, 459,
  1201

\bibitem[{{Ogilvie} \& {Lin}(2004)}]{ogilvie+lin2004}
{Ogilvie}, G.~I. \& {Lin}, D.~N.~C. 2004, \apj, 610, 477

\bibitem[{{Ouazzani} {et~al.}(2012){Ouazzani}, {Dupret}, \&
  {Reese}}]{Ouazzani2012}
{Ouazzani}, R.~M., {Dupret}, M.~A., \& {Reese}, D.~R. 2012, \aap, 547, A75

\bibitem[{{Ouazzani} {et~al.}(2020){Ouazzani}, {Ligni{\`e}res}, {Dupret},
  {Salmon}, {Ballot}, {Christophe}, \& {Takata}}]{Ouazzani2020Gamma}
{Ouazzani}, R.~M., {Ligni{\`e}res}, F., {Dupret}, M.~A., {et~al.} 2020, \aap,
  640, A49

\bibitem[{{Ouazzani} {et~al.}(2019){Ouazzani}, {Marques}, {Goupil},
  {Christophe}, {Antoci}, {Salmon}, \& {Ballot}}]{ouazzani2019}
{Ouazzani}, R.~M., {Marques}, J.~P., {Goupil}, M.~J., {et~al.} 2019, \aap, 626,
  A121

\bibitem[{{Ouazzani} {et~al.}(2015){Ouazzani}, {Roxburgh}, \&
  {Dupret}}]{Ouazzani2015}
{Ouazzani}, R.~M., {Roxburgh}, I.~W., \& {Dupret}, M.~A. 2015, \aap, 579, A116

\bibitem[{{Ouazzani} {et~al.}(2017){Ouazzani}, {Salmon}, {Antoci}, {Bedding},
  {Murphy}, \& {Roxburgh}}]{ouazzani2017}
{Ouazzani}, R.-M., {Salmon}, S.~J.~A.~J., {Antoci}, V., {et~al.} 2017, \mnras,
  465, 2294

\bibitem[{{P{\'a}pics} {et~al.}(2015){P{\'a}pics}, {Tkachenko}, {Aerts}, {Van
  Reeth}, {De Smedt}, {Hillen}, {{\O}stensen}, \& {Moravveji}}]{Papics2015B}
{P{\'a}pics}, P.~I., {Tkachenko}, A., {Aerts}, C., {et~al.} 2015, \apjl, 803,
  L25

\bibitem[{{P{\'a}pics} {et~al.}(2017){P{\'a}pics}, {Tkachenko}, {Van Reeth},
  {Aerts}, {Moravveji}, {Van de Sande}, {De Smedt}, {Bloemen}, {Southworth},
  {Debosscher}, {Niemczura}, \& {Gameiro}}]{Papics2017}
{P{\'a}pics}, P.~I., {Tkachenko}, A., {Van Reeth}, T., {et~al.} 2017, \aap,
  598, A74

\bibitem[{{Park} {et~al.}(2021){Park}, {Prat}, {Mathis}, \&
  {Bugnet}}]{park2021}
{Park}, J., {Prat}, V., {Mathis}, S., \& {Bugnet}, L. 2021, \aap, 646, A64

\bibitem[{{Pedersen} {et~al.}(2021){Pedersen}, {Aerts}, {P{\'a}pics},
  {Michielsen}, {Gebruers}, {Rogers}, {Molenberghs}, {Burssens}, {Garcia}, \&
  {Bowman}}]{Pedersen2021B}
{Pedersen}, M.~G., {Aerts}, C., {P{\'a}pics}, P.~I., {et~al.} 2021, Nature
  Astronomy, 5, 715

\bibitem[{{Prat} {et~al.}(2016){Prat}, {Ligni{\`e}res}, \& {Ballot}}]{Prat2016}
{Prat}, V., {Ligni{\`e}res}, F., \& {Ballot}, J. 2016, \aap, 587, A110

\bibitem[{{Prat} {et~al.}(2018){Prat}, {Mathis}, {Augustson}, {Ligni{\`e}res},
  {Ballot}, {Alvan}, \& {Brun}}]{Prat2018}
{Prat}, V., {Mathis}, S., {Augustson}, K., {et~al.} 2018, \aap, 615, A106

\bibitem[{{Prat} {et~al.}(2019){Prat}, {Mathis}, {Buysschaert}, {Van Beeck},
  {Bowman}, {Aerts}, \& {Neiner}}]{prat2019}
{Prat}, V., {Mathis}, S., {Buysschaert}, B., {et~al.} 2019, \aap, 627, A64

\bibitem[{{Prat} {et~al.}(2017){Prat}, {Mathis}, {Ligni{\`e}res}, {Ballot}, \&
  {Culpin}}]{prat2017}
{Prat}, V., {Mathis}, S., {Ligni{\`e}res}, F., {Ballot}, J., \& {Culpin}, P.~M.
  2017, \aap, 598, A105

\bibitem[{{Prat} {et~al.}(2020){Prat}, {Mathis}, {Neiner}, {Van Beeck},
  {Bowman}, \& {Aerts}}]{prat2020}
{Prat}, V., {Mathis}, S., {Neiner}, C., {et~al.} 2020, \aap, 636, A100

\bibitem[{{Reese} {et~al.}(2006){Reese}, {Ligni{\`e}res}, \&
  {Rieutord}}]{reese2006}
{Reese}, D., {Ligni{\`e}res}, F., \& {Rieutord}, M. 2006, \aap, 455, 621

\bibitem[{{Reese} {et~al.}(2014){Reese}, {Lara}, \& {Rieutord}}]{Reese2014}
{Reese}, D.~R., {Lara}, F.~E., \& {Rieutord}, M. 2014, in Precision
  Asteroseismology, ed. J.~A. {Guzik}, W.~J. {Chaplin}, G.~{Handler}, \&
  A.~{Pigulski}, Vol. 301, 169--172

\bibitem[{{Reese} {et~al.}(2021){Reese}, {Mirouh}, {Espinosa Lara}, {Rieutord},
  \& {Putigny}}]{Reese2021}
{Reese}, D.~R., {Mirouh}, G.~M., {Espinosa Lara}, F., {Rieutord}, M., \&
  {Putigny}, B. 2021, \aap, 645, A46

\bibitem[{{Ricker} {et~al.}(2014){Ricker}, {Winn}, {Vanderspek}, {Latham},
  {Bakos}, {Bean}, {Berta-Thompson}, {Brown}, {Buchhave}, {Butler}, {Butler},
  {Chaplin}, {Charbonneau}, {Christensen-Dalsgaard}, {Clampin}, {Deming},
  {Doty}, {De Lee}, {Dressing}, {Dunham}, {Endl}, {Fressin}, {Ge}, {Henning},
  {Holman}, {Howard}, {Ida}, {Jenkins}, {Jernigan}, {Johnson}, {Kaltenegger},
  {Kawai}, {Kjeldsen}, {Laughlin}, {Levine}, {Lin}, {Lissauer}, {MacQueen},
  {Marcy}, {McCullough}, {Morton}, {Narita}, {Paegert}, {Palle}, {Pepe},
  {Pepper}, {Quirrenbach}, {Rinehart}, {Sasselov}, {Sato}, {Seager},
  {Sozzetti}, {Stassun}, {Sullivan}, {Szentgyorgyi}, {Torres}, {Udry}, \&
  {Villasenor}}]{ricker2014}
{Ricker}, G.~R., {Winn}, J.~N., {Vanderspek}, R., {et~al.} 2014, in Society of
  Photo-Optical Instrumentation Engineers (SPIE) Conference Series, Vol. 9143,
  Space Telescopes and Instrumentation 2014: Optical, Infrared, and Millimeter
  Wave, ed. J.~{Oschmann}, Jacobus~M., M.~{Clampin}, G.~G. {Fazio}, \& H.~A.
  {MacEwen}, 914320

\bibitem[{{Rieutord}(2006)}]{Rieutord2006}
{Rieutord}, M. 2006, \aap, 451, 1025

\bibitem[{{Rieutord} {et~al.}(2005){Rieutord}, {Dintrans}, {Ligni{\`e}res},
  {Corbard}, \& {Pichon}}]{rieutord2005}
{Rieutord}, M., {Dintrans}, B., {Ligni{\`e}res}, F., {Corbard}, T., \&
  {Pichon}, B. 2005, in SF2A-2005: Semaine de l'Astrophysique Francaise, ed.
  F.~{Casoli}, T.~{Contini}, J.~M. {Hameury}, \& L.~{Pagani}, 759

\bibitem[{{Rieutord} {et~al.}(2016){Rieutord}, {Espinosa Lara}, \&
  {Putigny}}]{Rieutord2016}
{Rieutord}, M., {Espinosa Lara}, F., \& {Putigny}, B. 2016, Journal of
  Computational Physics, 318, 277

\bibitem[{Rogers(2015)}]{rogers2015}
Rogers, T.~M. 2015, The Astrophysical Journal, 815, L30

\bibitem[{{Rogers} {et~al.}(2013){Rogers}, {Lin}, {McElwaine}, \&
  {Lau}}]{Rogers2013}
{Rogers}, T.~M., {Lin}, D.~N.~C., {McElwaine}, J.~N., \& {Lau}, H.~H.~B. 2013,
  \apj, 772, 21

\bibitem[{{Roxburgh}(2006)}]{Roxburgh2006}
{Roxburgh}, I.~W. 2006, \aap, 454, 883

\bibitem[{{Saio} {et~al.}(2018){Saio}, {Bedding}, {Kurtz}, {Murphy}, {Antoci},
  {Shibahashi}, {Li}, \& {Takata}}]{Saio2018Gamma}
{Saio}, H., {Bedding}, T.~R., {Kurtz}, D.~W., {et~al.} 2018, \mnras, 477, 2183

\bibitem[{{Saio} {et~al.}(2015){Saio}, {Kurtz}, {Takata}, {Shibahashi},
  {Murphy}, {Sekii}, \& {Bedding}}]{Saio2015A}
{Saio}, H., {Kurtz}, D.~W., {Takata}, M., {et~al.} 2015, \mnras, 447, 3264

\bibitem[{{Saio} {et~al.}(2021){Saio}, {Takata}, {Lee}, {Li}, \& {Van
  Reeth}}]{Saio2021Gamma}
{Saio}, H., {Takata}, M., {Lee}, U., {Li}, G., \& {Van Reeth}, T. 2021, \mnras,
  502, 5856

\bibitem[{{Schmid} \& {Aerts}(2016)}]{Schmid2016F}
{Schmid}, V.~S. \& {Aerts}, C. 2016, \aap, 592, A116

\bibitem[{{Sowicka} {et~al.}(2017){Sowicka}, {Handler}, {D{\k{e}}bski},
  {Jones}, {Van de Sande}, \& {P{\'a}pics}}]{Sowicka2017F}
{Sowicka}, P., {Handler}, G., {D{\k{e}}bski}, B., {et~al.} 2017, \mnras, 467,
  4663

\bibitem[{{Szewczuk} \& {Daszy{\'n}ska-Daszkiewicz}(2018)}]{Szewczuk2018}
{Szewczuk}, W. \& {Daszy{\'n}ska-Daszkiewicz}, J. 2018, \mnras, 478, 2243

\bibitem[{{Szewczuk} {et~al.}(2021){Szewczuk}, {Walczak}, \&
  {Daszy{\'n}ska-Daszkiewicz}}]{Szewczuk2021B}
{Szewczuk}, W., {Walczak}, P., \& {Daszy{\'n}ska-Daszkiewicz}, J. 2021, \mnras,
  503, 5894

\bibitem[{{Tayar} {et~al.}(2019){Tayar}, {Beck}, {Pinsonneault}, {Garc{\'\i}a},
  \& {Mathur}}]{Tayar2019RG}
{Tayar}, J., {Beck}, P.~G., {Pinsonneault}, M.~H., {Garc{\'\i}a}, R.~A., \&
  {Mathur}, S. 2019, \apj, 887, 203

\bibitem[{{Triana} {et~al.}(2017){Triana}, {Corsaro}, {De Ridder}, {Bonanno},
  {P{\'e}rez Hern{\'a}ndez}, \& {Garc{\'\i}a}}]{Triana2017RG}
{Triana}, S.~A., {Corsaro}, E., {De Ridder}, J., {et~al.} 2017, \aap, 602, A62

\bibitem[{{Triana} {et~al.}(2015){Triana}, {Moravveji}, {P{\'a}pics}, {Aerts},
  {Kawaler}, \& {Christensen-Dalsgaard}}]{Triana2015B}
{Triana}, S.~A., {Moravveji}, E., {P{\'a}pics}, P.~I., {et~al.} 2015, \apj,
  810, 16

\bibitem[{{Unno} {et~al.}(1989){Unno}, {Osaki}, {Ando}, {Saio}, \&
  {Shibahashi}}]{unno1989}
{Unno}, W., {Osaki}, Y., {Ando}, H., {Saio}, H., \& {Shibahashi}, H. 1989,
  {Nonradial oscillations of stars} (University of Tokyo Press)

\bibitem[{{Van Beeck} {et~al.}(2020){Van Beeck}, {Prat}, {Van Reeth}, {Mathis},
  {Bowman}, {Neiner}, \& {Aerts}}]{VanBeeck2020}
{Van Beeck}, J., {Prat}, V., {Van Reeth}, T., {et~al.} 2020, \aap, 638, A149

\bibitem[{{Van Reeth} {et~al.}(2018){Van Reeth}, {Mombarg, J. S. G.}, {Mathis,
  S.}, {Tkachenko, A.}, {Fuller, J.}, {Bowman, D. M.}, {Buysschaert, B.},
  {Johnston, C.}, {Garc\'{\i}a Hern\'andez, A.}, {Goldstein, J.}, {Townsend, R.
  H. D.}, \& {Aerts, C.}}]{vanreeth2018}
{Van Reeth}, T., {Mombarg, J. S. G.}, {Mathis, S.}, {et~al.} 2018, A\&A, 618,
  A24

\bibitem[{{Van Reeth} {et~al.}(2016){Van Reeth}, {Tkachenko}, \&
  {Aerts}}]{vanreeth2016}
{Van Reeth}, T., {Tkachenko}, A., \& {Aerts}, C. 2016, A\&A, 593, A120

\bibitem[{{Van Reeth} {et~al.}(2015){Van Reeth}, {Tkachenko}, {Aerts},
  {P{\'a}pics}, {Degroote}, {Debosscher}, {Zwintz}, {Bloemen}, {De Smedt},
  {Hrudkova}, {Raskin}, \& {Van Winckel}}]{vanreeth2015}
{Van Reeth}, T., {Tkachenko}, A., {Aerts}, C., {et~al.} 2015, \aap, 574, A17

\bibitem[{Wang {et~al.}(2016)Wang, Boyd, \& Akmaev}]{wang2016}
Wang, H., Boyd, J.~P., \& Akmaev, R.~A. 2016, Geoscientific Model Development,
  9, 1477

\bibitem[{{Zahn}(1992)}]{zahn1992}
{Zahn}, J.~P. 1992, \aap, 265, 115

\end{thebibliography}
\end{document}